\documentclass[superscriptaddress,twocolumn,notitlepage,showkeys,showpacs,longbibliography]{revtex4-1}
\usepackage[utf8]{inputenc}
\usepackage{stmaryrd}
\usepackage{amsmath}
\usepackage{amssymb}
\usepackage{graphicx}
\usepackage{dcolumn}
\usepackage{bm}
\usepackage{multirow}
\usepackage[colorlinks,linkcolor=blue,citecolor=blue,urlcolor=blue]{hyperref}
\usepackage{color}
\usepackage{ulem}
\usepackage{comment}
\usepackage{booktabs}

\newcommand{\THX}[1]{{\color{black} #1}}

\begin{document}
\title{
Competing Lattice Instability and Magnetism on the Surface of Kagome Metals
}
\author{Hengxin Tan}
 \email{hengxin.tan@weizmann.ac.il}
\author{Binghai Yan}
 \email{binghai.yan@weizmann.ac.il}
\affiliation{Department of Condensed Matter Physics, Weizmann Institute of Science, Rehovot 7610001, Israel}
\date{\today}

\begin{abstract}
\THX{Only a few magnetic kagome materials exhibit lattice instabilities among the large kagome material array. 
In this work,  we find that kagome magnets $R$Mn$_6$Sn$_6$ ($R$ = rare-earth elements) and their non-magnetic counterparts $R$V$_6$Sn$_6$ exhibit intriguing interplay between magnetism and lattice dynamics on their surfaces. 
Notably, $R$V$_6$Sn$_6$ surfaces terminated by kagome layers demonstrate pronounced lattice instabilities, manifesting as $\sqrt{3} \times \sqrt{3}$ and $2 \times 2$ surface charge orders (SCOs). These instabilities are absent on corresponding magnetic materials $R$Mn$_6$Sn$_6$. Here, the SCO is suppressed by magnetism. Otherwise,
surface distortions would significantly reduce the spin polarization, elevating the energy via Hund's rule. Thus,  SCOs are energetically unfavorable on magnetic surfaces.
The competition of magnetism and lattice instability is further substantiated by observing SCOs on V-substituted $R$Mn$_6$Sn$_6$ kagome surfaces, contrasting with their absence on Mn-substituted $R$V$_6$Sn$_6$ surfaces.
Our findings reveal unexpected surface instability and profound spin-lattice coupling in these kagome magnets, highlighting the complex dynamics hidden within magnetic materials.}
\end{abstract}

\maketitle

\section{Introduction}
\THX{Kagome materials, characterized by a corner-sharing triangular sublattice, are a fertile playground for various correlated electronic orders \cite{wen2010interaction,Kiesel2012sublattice,wang2013Competing,Kiesel2013unconventional}.
The discovery of charge density wave (CDW) in non-magnetic kagome superconductors \cite{Ortiz2020Z2,jiang2021unconventional,tan2021charge,Liang2021three,chen2021roton} sparked immediate interest because of its close relationship with the potential time-reversal symmetry breaking and superconductivity. In contrast, magnetic kagome materials showing CDW are rare, primarily due to the lack of lattice instability in most systems and the competition between magnetism and other low-energy scale orders.

The recent discovery of CDW in the magnetic kagome metal FeGe \cite{teng2022discovery} highlights the intricate interplay between magnetism and lattice instability \cite{teng2023magnetism,miao2023signature,wang2023enhanced,chen2023longranged}, enabling the exploration of intertwining between these orders. This FeGe-type structure is derived by excluding the $R$ element from the extensively studied magnetic kagome compounds $R$Mn$_6$Sn$_6$ ($R$ = Y, Gd-Lu) \cite{Ghimire2020Competing,Riberolles2022low,yin2020quantum,zhang2020Topological,li2021dirac,xu2022topological,li2022manipulation,Asaba2020anomalous,ma2021rare,Gao2021anomalous,zhang2022exchange}, where no charge instability has been reported. This structural resemblance raises the possibility of an underlying lattice instability in $R$Mn$_6$Sn$_6$, potentially suppressed by magnetism and observable when magnetism is diminished. The observation of CDW transition in the non-magnetic ScV$_6$Sn$_6$ \cite{Arachchige2022charge,tan2023PRL,cao2023competing,Korshunov2023softening}, as opposed to its magnetic counterpart ScMn$_6$Sn$_6$ without CDW, suggests such a potential. However, the absence of CDW in other non-magnetic $R$V$_6$Sn$_6$ variants underscores the complexity of these materials. This complexity prompts questions about the existence of competition between lattice instability and magnetism and its manifestation in the wider range. Our recent Scanning Tunneling Microscopy (STM) studies reveal significant distortions on the surface of ScV$_6$Sn$_6$ \cite{cheng2023nanoscale}, hinting at the possibility of observing this competition in a decreased dimension.}

\THX{In this study, we elucidate the interplay between lattice instabilities and magnetism in surfaces of $R$V$_6$Sn$_6$ and $R$Mn$_6$Sn$_6$.} We reveal that the kagome layer on $R$V$_6$Sn$_6$ surfaces inherently exhibits charge instabilities, manifesting as $\sqrt{3} \times \sqrt{3}$ and $2 \times 2$ surface charge orders (SCOs), driven by phonon softening and characterized by distinct V trimer and clover-like distortions.
\THX{Introducing magnetic Mn in place of non-magnetic V on these kagome surfaces leads to a significant weakening of Mn spin polarization due to surface distortions, increasing the magnetic energy via Hund's rule. This energy increase suppresses SCOs, further evidenced by the absence of SCOs in magnetic $R$Mn$_6$Sn$_6$. Remarkably, substituting Mn with V on $R$Mn$_6$Sn$_6$ kagome surfaces allows SCOs to reappear, underlining a potent competition between surface lattice instability and magnetism and highlighting a pathway to potential high-temperature SCOs.
Our findings unveil complex dynamics in surfaces of kagome magnets and beyond.}


\section{Results and discussions}
\begin{figure}[tbp]
\includegraphics[width=0.95\linewidth]{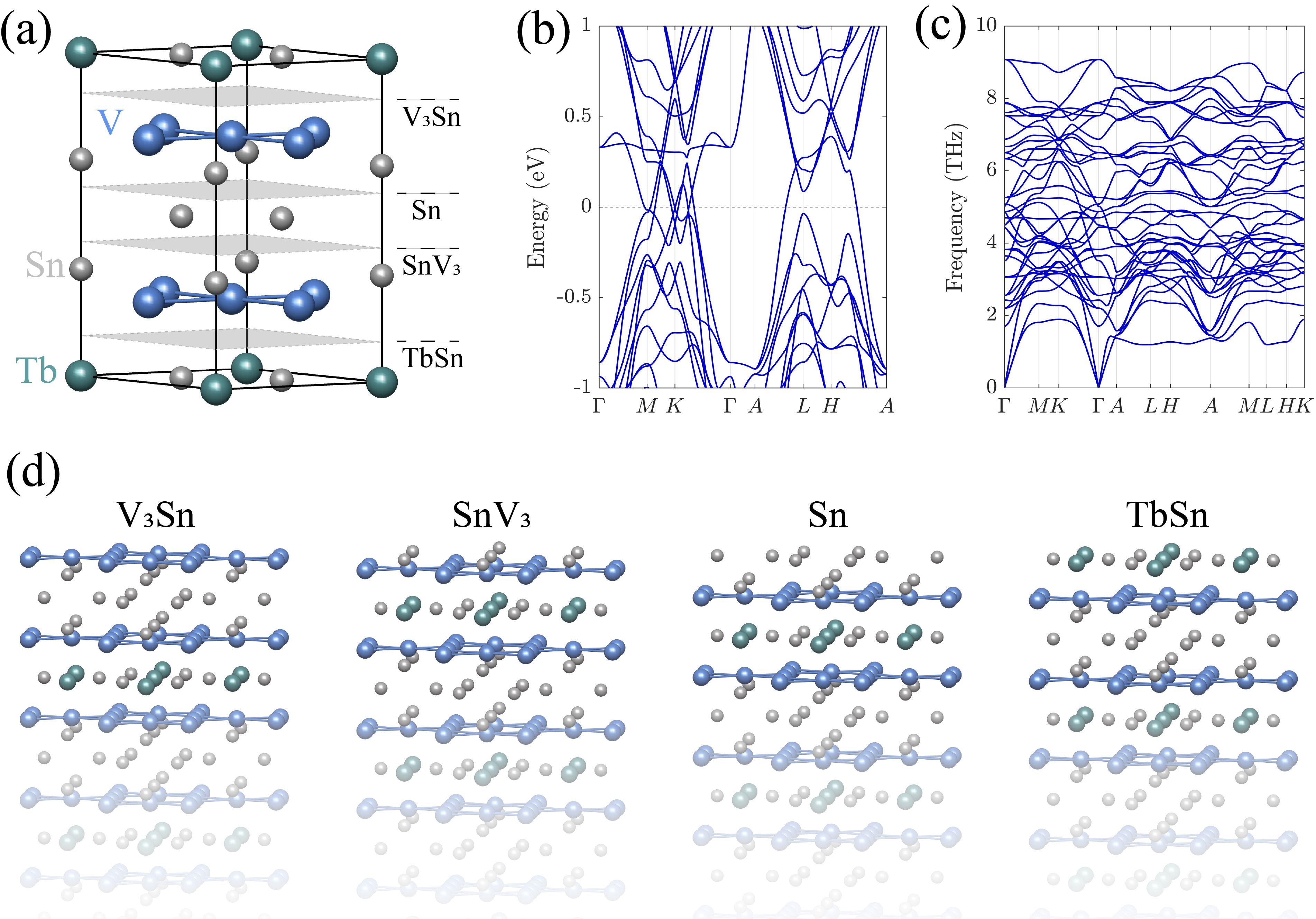}
\caption{\label{Fig1} \textbf{Bulk properties of TbV$_6$Sn$_6$ and surface configurations}. (a) The bulk crystal structure of TbV$_6$Sn$_6$. Grey planes stand for cleaving planes which generate four different terminations as labeled.
(b) Bulk band structure without spin-orbital coupling. (c) Bulk phonon dispersion.
(d) Four free-cleaved terminations. Both V$_3$Sn and SnV$_3$ are referred to as kagome surfaces. Notice that the surface V-kagome sublattice in the V$_3$Sn termination goes spontaneously below the highest Sn sublattice after surface relaxation.
}
\end{figure}

\textbf{Bulk properties of $R$V$_6$Sn$_6$}.
The crystalline structure of $R$V$_6$Sn$_6$ comprises a Sn honeycomb layer, a $R$Sn triangular layer, and two V-Sn kagome layers in a buckled configuration, as shown in Fig. \ref{Fig1}(a).
Prior experimental observations have identified a complex ordering of the $R$ 4$f$ magnetic moments below $\sim$4 K \cite{Pokharel2022highly,Rosenberg2022uniaxial,zhang2022electronic,lee2022anisotropic}. ARPES analysis has revealed the minimal influence of the 4$f$ magnetism on the low energy physics \cite{peng2021realizing,hu2022tunable,Sante2023flat}.
Our calculations demonstrate robust SCOs not only in the ferromagnetic TbV$_6$Sn$_6$ but also in paramagnetic GdV$_6$Sn$_6$ and nonmagnetic YV$_6$Sn$_6$. 
Consequently, we will omit discussions regarding $f$-electron magnetism.
Given the similarity of results across all $R$V$_6$Sn$_6$ compounds, we will primarily illustrate our findings with TbV$_6$Sn$_6$.

The bulk band structure of TbV$_6$Sn$_6$ is depicted in Fig. \ref{Fig1}(b), characterized by typical kagome band features, such as van Hove singularities at $M$/$L$, Dirac points at $K$/$H$, and a flat band at approximately 0.3 eV. Notice that all $R$V$_6$Sn$_6$ materials exhibit similar band structures \cite{Pokharel2021electronic,peng2021realizing,hu2022tunable,tan2023PRL,ding2023kagome}.
In Fig. \ref{Fig1}(c), the phonon spectrum of TbV$_6$Sn$_6$ confirms its bulk dynamical stability by showing the absence of imaginary phonon modes.
This absence extends to all $R$V$_6$Sn$_6$ compounds (except for ScV$_6$Sn$_6$ \cite{tan2023PRL}), indicating all $R$V$_6$Sn$_6$ compounds (except for Sc) are stable in their bulk forms. It's worth noting that, among these stable compounds, LuV$_6$Sn$_6$ exhibits the softest phonon branch on the $k_z$ plane, which may suggest its susceptibility to structural phase transitions under external influences like strain (see Supplementary II.A \cite{SM} for additional details).

\begin{figure}[tbp]
\includegraphics[width=0.95\linewidth]{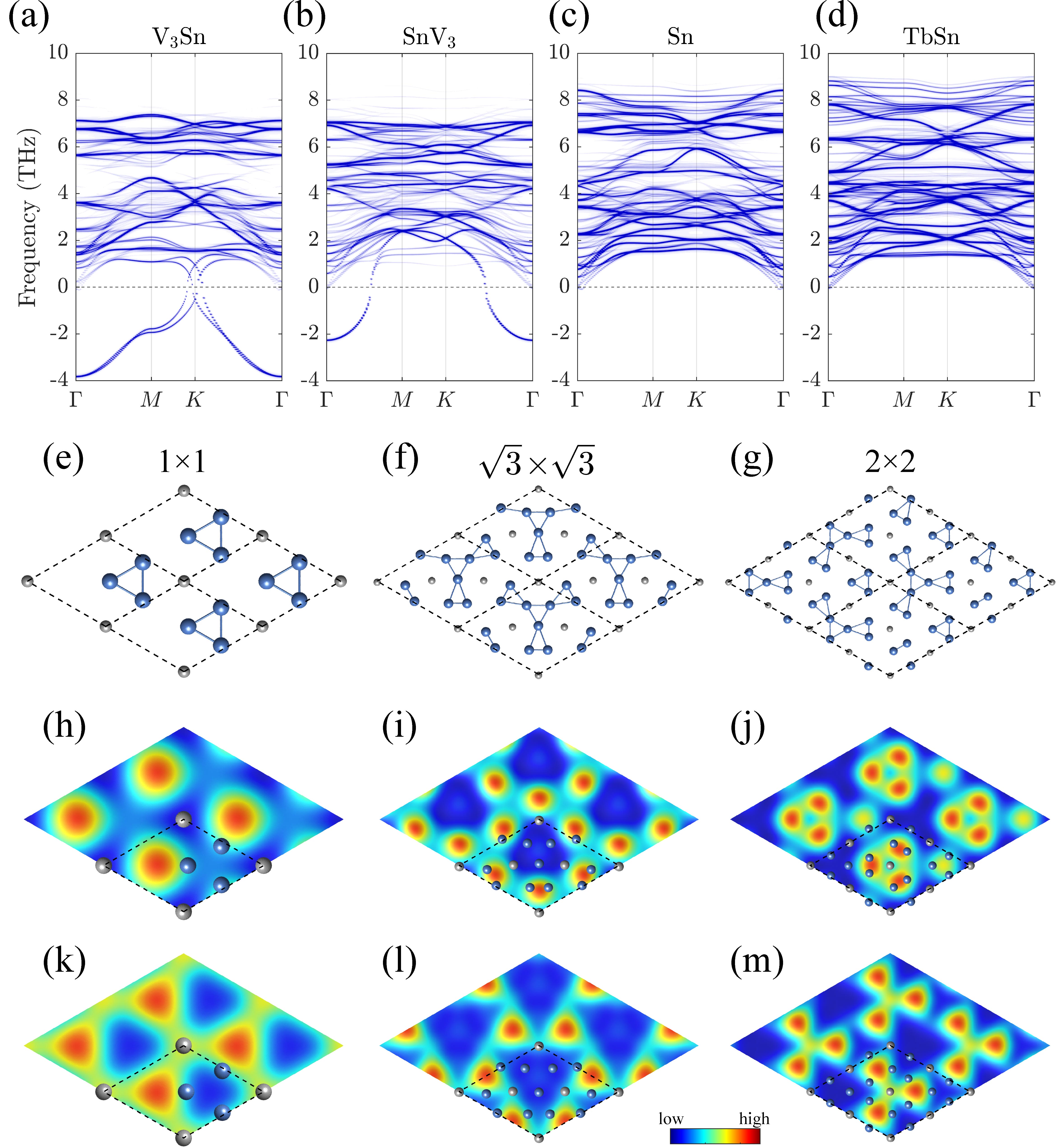}
\caption{\label{Fig2} \textbf{SCOs of TbV$_6$Sn$_6$}. (a)-(d) display surface phonon spectra of different terminations derived by projecting the phonon dispersions from four-unit-cell thick slabs onto the surface layers. Negative frequencies signify imaginary phonon modes.
(e)-(g) illustrate the $\Gamma$, $K$, $M$ imaginary phonon mode-driven distortions in the V$_3$Sn surface: (e) $1 \times 1$ V trimerization, (f) $\sqrt 3 \times \sqrt 3$ SCO (clover-like pattern), (g) $2 \times 2$ SCO (mixed V trimer and clover).
For clarity, only the topmost V and Sn sublattices are depicted in (e)-(g) for simplicity, with dashed diamonds denoting the smallest periodic cells.
The $\sqrt 3 \times \sqrt 3$ and $2 \times 2$ SCOs are also stable in the SnV$_3$ surface despite the absence of imaginary phonons at $M$ and $K$ in (b).
(h)-(j) depict STM images for $1 \times 1$, $\sqrt 3 \times \sqrt 3$ and $2 \times 2$ SCOs, respectively, in the V$_3$Sn surface at \THX{charge neutral point}.
(k)-(m) present analogous STM images for the SnV$_3$ surface.
}
\end{figure}

\textbf{SCOs of $R$V$_6$Sn$_6$}.
The layered structure of $R$V$_6$Sn$_6$ promises four different terminations, i.e., the $R$Sn triangular surface, Sn honeycomb surface, and two kagome surfaces, V$_3$Sn and SnV$_3$, as shown in Fig. \ref{Fig1}(a)\&(d),
These different surfaces have been observed in various experiments \cite{yin2020quantum,cheng2023nanoscale,hu2022tunable,peng2021realizing,Rosenberg2022uniaxial,li2022manipulation}.
The primary differentiation between the two kagome surfaces arises from the respective chemical configurations underneath. In the V$_3$Sn termination, the V-kagome sublattice is positioned above the closely associated Sn sublattice from the same buckled V-Sn layer, with the subsequent layer being the Sn honeycomb layer \cite{note1}.
In contrast, the SnV$_3$ termination situates the V-kagome sublattice below the closely associated Sn sublattice, with the next layer the $R$Sn triangular layer. Both terminations show similar SCOs and are collectively called kagome surfaces within this work.

Figures \ref{Fig2}(a)-(d) display the surface phonon band structures of the four different terminations of TbV$_6$Sn$_6$, obtained from thick slab model calculations.
Notably, the V$_3$Sn surface exhibits imaginary phonon modes across the entire surface Brillouin zone [Fig. \ref{Fig2}(a)].
The interpretation of these modes unfolds as follows:
i) The imaginary phonon at $\Gamma$ results in a $1\times 1$ surface modulation, primarily involving the surface V layer trimerization [Fig. \ref{Fig2}(e)]. This modulation reduces the surface's rotational symmetry from 6-fold to 3-fold while preserving the translational symmetry. However, such a trimerized surface is marked by a significant imaginary phonon at the $K$ point (see Supplementary II.C \cite{SM}), indicating its instability.
ii) The imaginary phonon at $K$ induces a $\sqrt 3 \times \sqrt 3$ SCO, as depicted in Fig. \ref{Fig2}(f). In this phase with broken translational symmetry, the surface V-kagome layer transforms into a clover-like pattern, resulting in a total energy decrease of about 307 meV per surface primitive unit cell (u.c.), relative to the pristine surface (see Table \ref{Table1}).
iii) The imaginary phonon at $M$ drives a $2 \times 2$ SCO, combining V trimers and clovers, as illustrated in Fig. \ref{Fig2}(g). This phase exhibits a lower total energy of $-$394 meV/u.c.
The dynamical stability of these two SCOs is confirmed by the absence of imaginary phonon modes in their respective phonon spectra (see Supplementary II.C \cite{SM}).
Typical scanning tunneling microscopy (STM) images at the \THX{charge neutral point} are presented in Fig. \ref{Fig2}(h)-(j) with additional results available in Supplementary II.H \cite{SM}.

In Fig. \ref{Fig2}(b), the phonon spectrum of the SnV$_3$ termination reveals exclusively imaginary phonons at $\Gamma$ point. The $\Gamma$ imaginary phonon instigates the 1$\times 1$ surface modulation with V trimers as illustrated in Fig. \ref{Fig2}(e), which reduces the total energy by $-$17 meV/u.c.. Intriguingly, such a V trimerized SnV$_3$ surface shows a substantial imaginary phonon at $K$, akin to the scenario in V$_3$Sn, as detailed in Supplementary II.C \cite{SM}. Furthermore, the $\sqrt 3 \times \sqrt 3$ and $2\times 2$ SCOs displayed in Fig. \ref{Fig2}(f)\&(g) elicit more pronounced reductions in the total energy of the SnV$_3$ surface, amounting to $-$26 and $-$21 meV/u.c., respectively. An insightful deduction from the phonon band structure is that an imaginary phonon mode at a reciprocal space momentum generally implies the presence of a stable structure modulated by the same momentum. The SnV$_3$ surface, however, presents a paradoxical case by revealing the absence of a stable structure modulated by the $\Gamma$ imaginary phonon and the unexpected emergence of energy-favored structures driven by the non-imaginary phonons at $M/K$. This peculiarity might arise from the effective screening of the long-range $\Gamma$ imaginary phonon to the short-range $M/K$ imaginary phonons, which manifests only after the elimination of the $\Gamma$ imaginary phonon. Notably, this $\Gamma$ imaginary phonon screening effect may also apply to the V$_3$Sn surface, exhibiting strong imaginary phonons at $\Gamma$ and weak ones at $M/K$. STM images of SCOs for the SnV$_3$ surface at the \THX{charge neutral point} are presented in Fig. \ref{Fig2}(k)-(m), with additional details available in Supplementary II.H \cite{SM}.

The phonon spectra of the Sn and TbSn terminations in Fig. \ref{Fig2}(c)\&(d) notably lack any soft modes. The above-established SCOs in kagome terminations are also confirmed to be unstable on both surfaces. It is crucial to emphasize that these two surfaces can be achieved by overlaying one Sn (TbSn) layer atop the SnV$_3$ (V$_3$Sn) surface. Consequently, introducing additional layers effectively mitigates the instabilities observed in the kagome surfaces. This comparison underscores the pivotal role of exposing the kagome layer at the surface in attaining SCOs in $R$V$_6$Sn$_6$. Furthermore, it is worth noting that there are numerous flat phonon bands evident across all surfaces, which could potentially influence the phonon dynamics, as discussed in \cite{hu2023kagome}.

\begin{table}
\centering
\caption{\label{Table1} \textbf{Total energy of SCOs on kagome terminations of $R$V$_6$Sn$_6$}. The energy, with respect to the pristine termination, is in the unit of meV/u.c. (u.c. = surface primitive unit cell).
Notice that under-waved $1\times 1$ distortions are confirmed to show imaginary phonon modes at $K$ point.
The $2\times 2$ SCO in the SnV$_3$ termination of other $R$ materials (except for Gd and Tb) and the $\sqrt 3 \times \sqrt 3$ SCO in the SnV$_3$ termination of ScV$_6$Sn$_6$ go back to the $1\times 1$ surface V trimerization after structural relaxation, indicating their instability.
Sc and Sc$^*$ stand for the bulk pristine and $\sqrt 3 \times \sqrt 3 \times 3$ CDW phase, respectively.
}
\renewcommand\arraystretch{1.2}
\begin{ruledtabular}
\begin{tabular}{ccccccccccc}
  \multirow{2}{*}{$R$} & & \multicolumn{3}{c}{V$_3$Sn}  & & \multicolumn{3}{c}{SnV$_3$} \\
  \cline{3-5}
  \cline{7-9}
  \specialrule{0em}{1pt}{1pt}
   && $\sqrt 3 \times \sqrt 3$ & $2\times 2$ & $1\times 1$ &  & $\sqrt 3 \times \sqrt 3$ & $2\times 2$ & $1\times 1$ \\
  \hline
    Gd  && $-$318 & $-$402 & $-$214  & & $-$36 & $-$33 & $-$26 \\
    Tb  && $-$307 & $-$394 & \uwave{$-$212}  & & $-$26 & $-$21 & \uwave{$-$17} \\
    Dy  && $-$304 & $-$392 & $-$204  & & $-$31 & Unstable & $-$25 \\
    Ho  && $-$300 & $-$391 & $-$200  & & $-$28 & Unstable & $-$23 \\
    Er  && $-$297 & $-$388 & $-$196  & & $-$25 & Unstable & $-$22 \\
    Tm  && $-$294 & $-$386 & $-$192  & & $-$22 & Unstable & $-$21 \\
    Lu  && $-$289 & $-$383 & $-$186  & & $-$17 & Unstable & $-$16 \\
    Y   && $-$305 & $-$393 & \uwave{$-$205}  & & $-$33 & Unstable & \uwave{$-$27} \\
   \specialrule{0em}{2pt}{2pt}
    Sc  && $-$251 & $-$359 & $-$155  & & Unstable & Unstable & $-$4\\
    Sc$^*$  && $-$94 &   &   & & Unstable &  & \\
\end{tabular}
\end{ruledtabular}
\end{table}

We apply the above SCOs to the kagome terminations of all other $R$V$_6$Sn$_6$.
After adequate surface relaxation, these SCOs consistently reduce the total energy, as summarized in Table \ref{Table1}.
Similar to the Tb compound, the $2\times 2$ SCO consistently exhibits a lower total energy in the V$_3$Sn termination, while the $\sqrt 3 \times \sqrt 3$ SCO attains a lower total energy in the SnV$_3$ termination across all $R$ compounds.
Remarkably, while the $2\times 2$ SCO can be achieved in the SnV$_3$ termination of GdV$_6$Sn$_6$ following surface relaxation, the $2\times 2$ distorted SnV$_3$ surfaces of Y and Dy-Lu materials spontaneously revert to the $1 \times 1$ trimerized surface configuration after relaxation, effectively restoring the broken translational symmetry.
Phonon calculations additionally reveal that the $1\times 1$ trimerized kagome surfaces of YV$_6$Sn$_6$ exhibit a pronounced imaginary phonon at $K$ (resulting in the $\sqrt 3 \times \sqrt 3$ SCO), notwithstanding the pristine SnV$_3$ surface of the Y compound only manifesting an imaginary phonon at the $\Gamma$ point (elaborated in Supplementary II.D \cite{SM}).
Given the resemblance of these materials, we conjecture that the $1\times 1$ trimerized kagome surfaces exhibit instability across all $R$V$_6$Sn$_6$ compounds (except for Sc). However, their total energies are included in Table \ref{Table1} for reference and comparative analysis.
These SCOs are pure surface effects, which are robust against the slab thickness as comprehensively elucidated in Supplementary II.E \cite{SM}.

We have also extended our analysis of SCOs to the extensively studied ScV$_6$Sn$_6$.
In the pristine bulk phase, while two SCOs reduce the energy of the V$_3$Sn termination, they are unstable in the SnV$_3$ termination, and only the $1 \times 1$ surface trimerization reduces the total energy. This phenomenon may be attributed to the significantly higher in-plane chemical pressure in ScV$_6$Sn$_6$ (see Supplementary II.A \cite{SM}).
Furthermore, we applied the $\sqrt 3 \times \sqrt 3$ SCO to kagome terminations of ScV$_6$Sn$_6$ in bulk $\sqrt 3 \times \sqrt 3 \times 3$ CDW phase \cite{Arachchige2022charge}, and found it to be only stabilized in the V$_3$Sn termination (Table \ref{Table1}).

\THX{
\textbf{Suppression of SCO by magnetism}.
Now we explore the impact of magnetism on SCOs by substituting the outermost V kagome layer on TbV$_6$Sn$_6$ kagome terminations with a magnetic Mn kagome layer. Calculations indicate that the local magnetic moments of Mn form a ferromagnetic ordering. This magnetic substitution leads to the disappearance of SCOs, following sufficient surface relaxation. 
Indeed, such Mn-doped surfaces with frozen SCOs show much higher total energies (Table \ref{Table2}).
Moreover, phonon spectra analyses of these Mn-doped kagome surfaces reveal no lattice instabilities, as shown in Figs. \ref{Fig3}(a). The interplay between SCOs and magnetism is further elucidated by studying magnetic counterpart TbMn$_6$Sn$_6$. Namely, the absence of imaginary phonon modes on kagome surfaces Mn$_3$Sn and SnMn$_3$ indicates no SCOs (see Supplementary II.G \cite{SM}). Conversely, reintroducing V into kagome surfaces of TbMn$_6$Sn$_6$ stabilizes the SCOs. Notably, the cross substitution between V and Mn does not exert a noticeable strain effect, attributed to the similar effective radii of V and Mn ions (V 0.640 Å versus Mn 0.645 Å \cite{shannon1969effective}). Thus, the stabilization or destabilization of SCOs in these kagome surfaces is more closely related to magnetic effects rather than mechanical strain.
Notice that the $f$-electron magnetism of Tb does not affect SCOs.

\begin{figure}[tbp]
\includegraphics[width=0.95\linewidth]{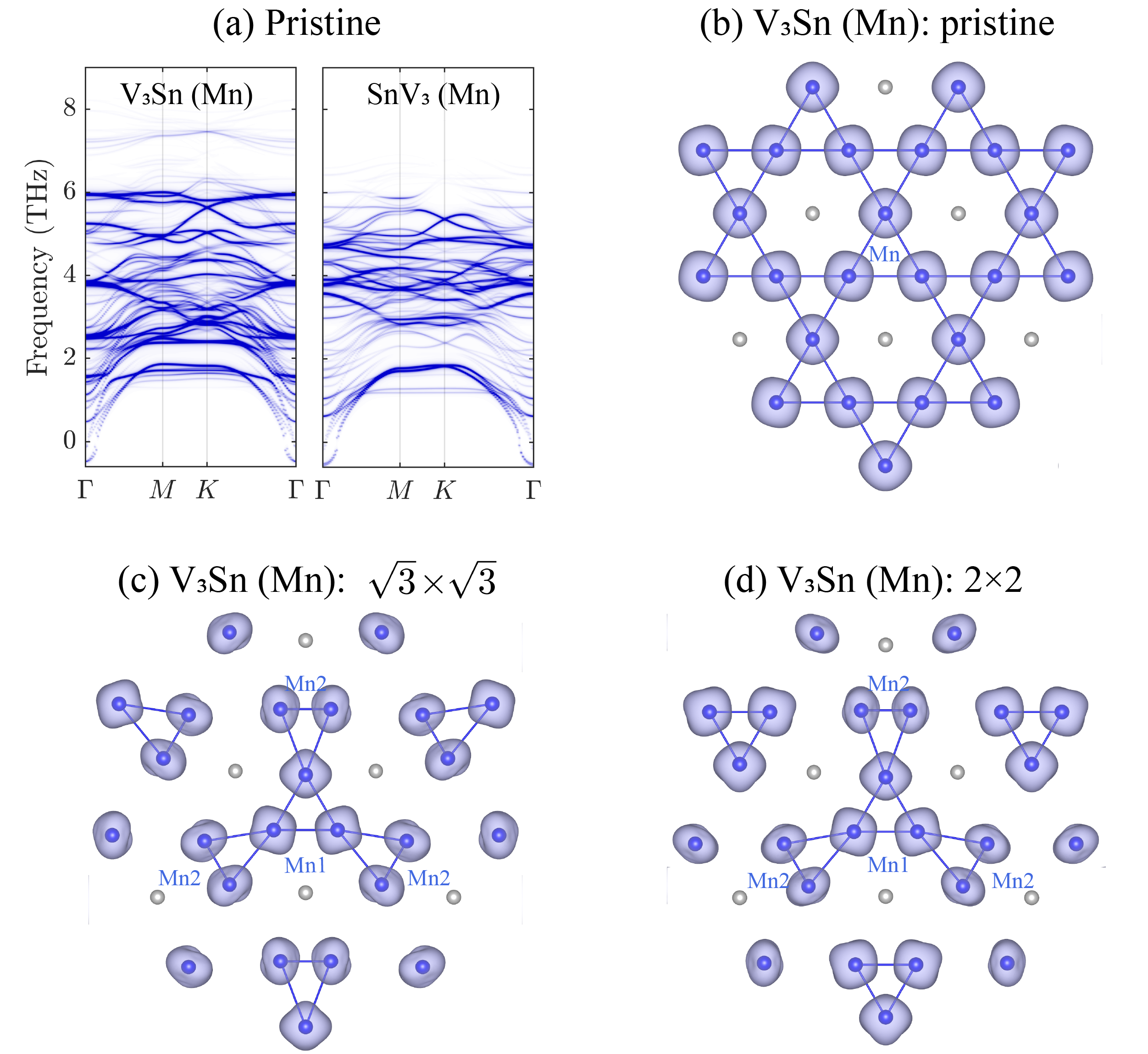}
\caption{\label{Fig3} \textbf{Phonon and magnetization density of Mn-doped kagome surfaces of TbV$_6$Sn$_6$}.
(a) shows phonon dispersions of two pristine surfaces, where Mn replaces the topmost V layer. The faint soft feature is caused by the calculation accuracy.
(b) displays the magnetization density of the Mn-replaced pristine V$_3$Sn surface, where the isosurface is 0.015 $e$/\AA$^3$.
(c) and (d) are similar to (b) but with frozen SCOs, where Mn2 atoms (blue) have a smaller magnetization density area, indicating their weaker spin polarization.
}
\end{figure}

\begin{table*}
\centering
\caption{\label{Table2} \textbf{Properties of Mn-doped kagome surfaces of TbV$_6$Sn$_6$.} The surface V kagome layer of TbV$_6$Sn$_6$ is replaced with Mn atoms where SCOs are frozen. Up and dn represent the spin-up and down channels, respectively. Mn atoms are labeled in Fig. \ref{Fig3}(b)-(d).
The first row of each surface shows the Bader charges (unit $e$);
the second row displays the magnetization $M$ (unit $\mu_B$), which is equivalent to the difference between spin up and down Bader charges;
the third row is the total energy $\delta E$ relative to the pristine surface (eV per pristine cell, p.c.).
The total charge of each Mn atom (spin-up + spin-down Bader charges) is nearly preserved by SCOs. SCOs transfer spin-up charges to the spin-down channel, leading to a decreased spin polarization (smaller magnetic moment).
}
\renewcommand\arraystretch{1.3}
\begin{ruledtabular}
\begin{tabular}{cccccccccccccccccccc}
  \multicolumn{2}{c}{Mn-replaced}  & & \multicolumn{2}{c}{Pristine}  && \multicolumn{4}{c}{$\sqrt 3\times \sqrt 3$ SCO} && \multicolumn{4}{c}{$2\times 2$ SCO} \\
  \cline{4-5}
  \cline{7-10}
  \cline{12-15}
  \multicolumn{2}{c}{kagome surface} & & Mn up & Mn dn & & Mn1 up & Mn1 dn & Mn2 up & Mn2 dn & & Mn1 up & Mn1 dn & Mn2 up & Mn2 dn \\
  \hline
  \multirow{3}{*}{V$_3$Sn} & charge ($e$) &&5.17 & 1.65 & & 4.84 & 2.02 & 4.34 & 2.53 & & 4.88 & 1.96 & 4.22 & 2.65 \\
    & $M$ ($\mu_B$)       &&\multicolumn{2}{c}{3.52} && \multicolumn{2}{c}{2.82} & \multicolumn{2}{c}{1.81} && \multicolumn{2}{c}{2.92} & \multicolumn{2}{c}{1.57} \\
    & $\delta E$ (eV/p.c.)  &&\multicolumn{2}{c}{0} && \multicolumn{4}{c}{1.23} && \multicolumn{4}{c}{1.29} \\
    \hline
    \multirow{3}{*}{SnV$_3$} & charge ($e$) &&5.11 & 1.75 & & 5.04 & 1.84 & 5.00 & 1.87 & & 5.02 & 1.86 & 4.99 & 1.88 \\
    & $M$ ($\mu_B$)       &&\multicolumn{2}{c}{3.36} && \multicolumn{2}{c}{3.20} & \multicolumn{2}{c}{3.13} && \multicolumn{2}{c}{3.16} & \multicolumn{2}{c}{3.11} \\
    & $\delta E$ (eV/p.c.) &&\multicolumn{2}{c}{0} && \multicolumn{4}{c}{0.38} && \multicolumn{4}{c}{0.54} \\
\end{tabular}
\end{ruledtabular}
\end{table*}

These observations highlight the competition between lattice instabilities and magnetism. A detailed Bader charge analysis of Mn atoms of the Mn-doped kagome surfaces of TbV$_6$Sn$_6$ with frozen SCOs is presented in Table \ref{Table2}. It shows that while the total charge of each Mn atom is preserved, SCOs transfer charges from the spin-up channel to the spin-down channel, weakening the spin polarization heavily.
For example, on the Mn-doped V$_3$Sn surface, Mn atoms have initial spin-up/down charges of 5.17$e$/1.65$e$, yielding a total charge of 6.82$e$ and magnetization of 3.52$\mu_B$. With the frozen $\sqrt 3 \times \sqrt 3$ SCO, spin-up charges adjust to 4.84$e$ (Mn1) and 4.34$e$ (Mn2, labeled in Fig. \ref{Fig3}(c)), and spin-down to 2.02$e$ (Mn1) and 2.53$e$ (Mn2), nearly preserving total charges (6.86$e$ for Mn1, 6.87$e$ for Mn2) but reducing magnetizations to 2.82$\mu_B$ (Mn1) and 1.81$\mu_B$ (Mn2).
Magnetization densities are exemplified with Mn-doped V$_3$Sn in Fig. \ref{Fig3}(b)-(d), where the Mn2 atoms have a much smaller magnetization density area under SCOs.
Similar behaviors are confirmed for all other surfaces and SCOs (changes are smaller in the SnV$_3$ surface because of its smaller SCO distortion amplitudes).
Results of TbMn$_6$Sn$_6$ kagome terminations are found in Supplementary II.G \cite{SM}.
According to Hund's rule which favors high-spin states in magnetic ions for maximum spin polarization, the reduction in spin polarization leads to increased magnetic energy, which outweighs the kinetic energy loss from structural distortions. As a result, the total energy of the surface is increased, mitigating surface distortions.
For TbV$_6$Sn$_6$ and V-doped TbMn$_6$Sn$_6$ kagome terminations, the elimination of surface magnetism ceases its antagonistic effect on SCOs, resulting in SCOs as previously discussed.}

This mechanism contrasts the scenario for the CDW formation of the kagome magnet FeGe. In FeGe, structural distortions lead to kinetic energy gains and increased spin polarization, reducing more magnetic energy \cite{miao2023signature,wang2023enhanced,chen2023longranged}. \THX{Moreover, spin-splitting shifts a von Hove singularity into the Fermi level in the spin minority band \cite{teng2023magnetism}, enhancing Fermi surface nesting and promoting charge instability.
The intricate interplay among spin, lattice, and charge degrees of freedom culminates in the CDW formation in FeGe and SCO suppression in $R$Mn$_6$Sn$_6$, illustrating the complex quantum ground states emergent from the intertwined dynamics of these variables.}

Now we address the potential experimental detection of the predicted SCOs.
SCOs significantly disrupt the surface kagome layers, obfuscating certain surface states (see Supplementary II.F \cite{SM}).
ARPES appears as the initial experimental technique to consider. Notably, a limited number of ARPES experiments have been conducted on kagome terminations of $R$V$_6$Sn$_6$ \cite{hu2022tunable,peng2021realizing,Rosenberg2022uniaxial,Sante2023flat,cheng2023nanoscale,ding2023kagome}. While experiments observe some distinctive bands, ambiguous states are also present whose origin remains challenging to identify. Consequently, the determination of the presence of SCOs in these experiments, as well as the characterization of their range (long-range or short-range), remains elusive.
We propose that microscope techniques such as STM, capable of detecting local atomic structures, could be an optimal approach for SCO detection.
Nevertheless, relatively few STM experiments have been conducted on $R$V$_6$Sn$_6$.
In the case of ScV$_6$Sn$_6$ within the bulk $\sqrt 3 \times \sqrt 3 \times 3$ CDW phase, recent STM experiments on kagome terminations observed a prominent $\sqrt 3 \times \sqrt 3$ reconstruction \cite{cheng2023nanoscale}. We do not conclude it the $\sqrt 3 \times \sqrt 3$ SCO (Table \ref{Table1}) because the bulk CDW also exhibits a $\sqrt 3 \times \sqrt 3$ in-plane modulation. However, it's crucial to underscore that the bulk CDW of ScV$_6$Sn$_6$ exclusively involves Sc/Sn out-of-plane movements, with minimal distortion in the V kagome layer \cite{Arachchige2022charge,tan2023PRL}. This contrasts the STM experiment, underscoring the need for more comprehensive studies for $R$V$_6$Sn$_6$.

In the context of $R$Mn$_6$Sn$_6$, intriguing phenomena are anticipated. It is noteworthy to consider the possibility of observing SCOs in $R$Mn$_6$Sn$_6$ at high temperatures when the magnetic order ceases to exist. Confirmation of this phenomenon in experiments would establish $R$Mn$_6$Sn$_6$ as a rare class of symmetry-descending materials when temperature increases \cite{wu2024symmetry}. We recommend more STM experiments on $R$Mn$_6$Sn$_6$ at elevated temperatures to explore this intriguing prospect. Furthermore, the sizeable structural distortion accompanying SCOs might also suggest the potential for a significant surface magnetostriction \cite{Lee1955Magnetostriction} and piezomagnetic effect, which would benefit the application of antiferromagnetic.

\section{Conclusion}
\THX{In conclusion, we have revealed the pervasive $\sqrt{3} \times \sqrt{3}$ and $2 \times 2$ SCOs across kagome terminations in $R$V$_6$Sn$_6$, which are suppressed in $R$Mn$_6$Sn$_6$ by magnetism due to the profound competition between lattice instability and magnetism through Hund's rule. We call for additional experimental efforts to explore these fascinating surface phenomena. Our findings suggest the existence of undiscovered quantum states in general magnets, potentially accessible through the suppression of magnetism.}



\section{Acknowledgements} 
We thank Ilija Zeljkovic for fruitful discussions.
H.T. also thanks Yizhou Liu for helpful discussions.
B.Y. acknowledges the financial support by the European Research Council (ERC Consolidator Grant ``NonlinearTopo'', No. 815869) and the ISF - Personal Research Grant	(No. 2932/21) and the DFG (CRC 183, A02).






\begin{thebibliography}{53}%
\makeatletter
\providecommand \@ifxundefined [1]{%
 \@ifx{#1\undefined}
}%
\providecommand \@ifnum [1]{%
 \ifnum #1\expandafter \@firstoftwo
 \else \expandafter \@secondoftwo
 \fi
}%
\providecommand \@ifx [1]{%
 \ifx #1\expandafter \@firstoftwo
 \else \expandafter \@secondoftwo
 \fi
}%
\providecommand \natexlab [1]{#1}%
\providecommand \enquote  [1]{``#1''}%
\providecommand \bibnamefont  [1]{#1}%
\providecommand \bibfnamefont [1]{#1}%
\providecommand \citenamefont [1]{#1}%
\providecommand \href@noop [0]{\@secondoftwo}%
\providecommand \href [0]{\begingroup \@sanitize@url \@href}%
\providecommand \@href[1]{\@@startlink{#1}\@@href}%
\providecommand \@@href[1]{\endgroup#1\@@endlink}%
\providecommand \@sanitize@url [0]{\catcode `\\12\catcode `\$12\catcode
  `\&12\catcode `\#12\catcode `\^12\catcode `\_12\catcode `\%12\relax}%
\providecommand \@@startlink[1]{}%
\providecommand \@@endlink[0]{}%
\providecommand \url  [0]{\begingroup\@sanitize@url \@url }%
\providecommand \@url [1]{\endgroup\@href {#1}{\urlprefix }}%
\providecommand \urlprefix  [0]{URL }%
\providecommand \Eprint [0]{\href }%
\providecommand \doibase [0]{http://dx.doi.org/}%
\providecommand \selectlanguage [0]{\@gobble}%
\providecommand \bibinfo  [0]{\@secondoftwo}%
\providecommand \bibfield  [0]{\@secondoftwo}%
\providecommand \translation [1]{[#1]}%
\providecommand \BibitemOpen [0]{}%
\providecommand \bibitemStop [0]{}%
\providecommand \bibitemNoStop [0]{.\EOS\space}%
\providecommand \EOS [0]{\spacefactor3000\relax}%
\providecommand \BibitemShut  [1]{\csname bibitem#1\endcsname}%
\let\auto@bib@innerbib\@empty
\bibitem [{\citenamefont {Wen}\ \emph {et~al.}(2010)\citenamefont {Wen},
  \citenamefont {R\"uegg}, \citenamefont {Wang},\ and\ \citenamefont
  {Fiete}}]{wen2010interaction}%
  \BibitemOpen
  \bibfield  {author} {\bibinfo {author} {\bibfnamefont {Jun}\ \bibnamefont
  {Wen}}, \bibinfo {author} {\bibfnamefont {Andreas}\ \bibnamefont {R\"uegg}},
  \bibinfo {author} {\bibfnamefont {C.-C.~Joseph}\ \bibnamefont {Wang}}, \ and\
  \bibinfo {author} {\bibfnamefont {Gregory~A.}\ \bibnamefont {Fiete}},\
  }\bibfield  {title} {\enquote {\bibinfo {title} {Interaction-driven
  topological insulators on the kagome and the decorated honeycomb lattices},}\
  }\href {\doibase 10.1103/PhysRevB.82.075125} {\bibfield  {journal} {\bibinfo
  {journal} {Phys. Rev. B}\ }\textbf {\bibinfo {volume} {82}},\ \bibinfo
  {pages} {075125} (\bibinfo {year} {2010})}\BibitemShut {NoStop}%
\bibitem [{\citenamefont {Kiesel}\ and\ \citenamefont
  {Thomale}(2012)}]{Kiesel2012sublattice}%
  \BibitemOpen
  \bibfield  {author} {\bibinfo {author} {\bibfnamefont {Maximilian~L.}\
  \bibnamefont {Kiesel}}\ and\ \bibinfo {author} {\bibfnamefont {Ronny}\
  \bibnamefont {Thomale}},\ }\bibfield  {title} {\enquote {\bibinfo {title}
  {Sublattice interference in the kagome hubbard model},}\ }\href {\doibase
  10.1103/PhysRevB.86.121105} {\bibfield  {journal} {\bibinfo  {journal} {Phys.
  Rev. B}\ }\textbf {\bibinfo {volume} {86}},\ \bibinfo {pages} {121105}
  (\bibinfo {year} {2012})}\BibitemShut {NoStop}%
\bibitem [{\citenamefont {Wang}\ \emph {et~al.}(2013)\citenamefont {Wang},
  \citenamefont {Li}, \citenamefont {Xiang},\ and\ \citenamefont
  {Wang}}]{wang2013Competing}%
  \BibitemOpen
  \bibfield  {author} {\bibinfo {author} {\bibfnamefont {Wan-Sheng}\
  \bibnamefont {Wang}}, \bibinfo {author} {\bibfnamefont {Zheng-Zhao}\
  \bibnamefont {Li}}, \bibinfo {author} {\bibfnamefont {Yuan-Yuan}\
  \bibnamefont {Xiang}}, \ and\ \bibinfo {author} {\bibfnamefont {Qiang-Hua}\
  \bibnamefont {Wang}},\ }\bibfield  {title} {\enquote {\bibinfo {title}
  {Competing electronic orders on kagome lattices at van hove filling},}\
  }\href {\doibase 10.1103/PhysRevB.87.115135} {\bibfield  {journal} {\bibinfo
  {journal} {Phys. Rev. B}\ }\textbf {\bibinfo {volume} {87}},\ \bibinfo
  {pages} {115135} (\bibinfo {year} {2013})}\BibitemShut {NoStop}%
\bibitem [{\citenamefont {Kiesel}\ \emph {et~al.}(2013)\citenamefont {Kiesel},
  \citenamefont {Platt},\ and\ \citenamefont
  {Thomale}}]{Kiesel2013unconventional}%
  \BibitemOpen
  \bibfield  {author} {\bibinfo {author} {\bibfnamefont {Maximilian~L.}\
  \bibnamefont {Kiesel}}, \bibinfo {author} {\bibfnamefont {Christian}\
  \bibnamefont {Platt}}, \ and\ \bibinfo {author} {\bibfnamefont {Ronny}\
  \bibnamefont {Thomale}},\ }\bibfield  {title} {\enquote {\bibinfo {title}
  {Unconventional fermi surface instabilities in the kagome hubbard model},}\
  }\href {\doibase 10.1103/PhysRevLett.110.126405} {\bibfield  {journal}
  {\bibinfo  {journal} {Phys. Rev. Lett.}\ }\textbf {\bibinfo {volume} {110}},\
  \bibinfo {pages} {126405} (\bibinfo {year} {2013})}\BibitemShut {NoStop}%
\bibitem [{\citenamefont {Ortiz}\ \emph {et~al.}(2020)\citenamefont {Ortiz},
  \citenamefont {Teicher}, \citenamefont {Hu}, \citenamefont {Zuo},
  \citenamefont {Sarte}, \citenamefont {Schueller}, \citenamefont {Abeykoon},
  \citenamefont {Krogstad}, \citenamefont {Rosenkranz}, \citenamefont {Osborn},
  \citenamefont {Seshadri}, \citenamefont {Balents}, \citenamefont {He},\ and\
  \citenamefont {Wilson}}]{Ortiz2020Z2}%
  \BibitemOpen
  \bibfield  {author} {\bibinfo {author} {\bibfnamefont {Brenden~R.}\
  \bibnamefont {Ortiz}}, \bibinfo {author} {\bibfnamefont {Samuel M.~L.}\
  \bibnamefont {Teicher}}, \bibinfo {author} {\bibfnamefont {Yong}\
  \bibnamefont {Hu}}, \bibinfo {author} {\bibfnamefont {Julia~L.}\ \bibnamefont
  {Zuo}}, \bibinfo {author} {\bibfnamefont {Paul~M.}\ \bibnamefont {Sarte}},
  \bibinfo {author} {\bibfnamefont {Emily~C.}\ \bibnamefont {Schueller}},
  \bibinfo {author} {\bibfnamefont {A.~M.~Milinda}\ \bibnamefont {Abeykoon}},
  \bibinfo {author} {\bibfnamefont {Matthew~J.}\ \bibnamefont {Krogstad}},
  \bibinfo {author} {\bibfnamefont {Stephan}\ \bibnamefont {Rosenkranz}},
  \bibinfo {author} {\bibfnamefont {Raymond}\ \bibnamefont {Osborn}}, \bibinfo
  {author} {\bibfnamefont {Ram}\ \bibnamefont {Seshadri}}, \bibinfo {author}
  {\bibfnamefont {Leon}\ \bibnamefont {Balents}}, \bibinfo {author}
  {\bibfnamefont {Junfeng}\ \bibnamefont {He}}, \ and\ \bibinfo {author}
  {\bibfnamefont {Stephen~D.}\ \bibnamefont {Wilson}},\ }\bibfield  {title}
  {\enquote {\bibinfo {title} {{$\mathrm{Cs}{\mathrm{V}}_{3}{\mathrm{Sb}}_{5}$:
  A ${\mathbb{Z}}_{2}$ Topological Kagome Metal with a Superconducting Ground
  State}},}\ }\href {\doibase 10.1103/PhysRevLett.125.247002} {\bibfield
  {journal} {\bibinfo  {journal} {Phys. Rev. Lett.}\ }\textbf {\bibinfo
  {volume} {125}},\ \bibinfo {pages} {247002} (\bibinfo {year}
  {2020})}\BibitemShut {NoStop}%
\bibitem [{\citenamefont {Jiang}\ \emph {et~al.}(2021)\citenamefont {Jiang},
  \citenamefont {Yin}, \citenamefont {Denner}, \citenamefont {Shumiya},
  \citenamefont {Ortiz}, \citenamefont {Xu}, \citenamefont {Guguchia},
  \citenamefont {He}, \citenamefont {Hossain}, \citenamefont {Liu} \emph
  {et~al.}}]{jiang2021unconventional}%
  \BibitemOpen
  \bibfield  {author} {\bibinfo {author} {\bibfnamefont {Yu-Xiao}\ \bibnamefont
  {Jiang}}, \bibinfo {author} {\bibfnamefont {Jia-Xin}\ \bibnamefont {Yin}},
  \bibinfo {author} {\bibfnamefont {M~Michael}\ \bibnamefont {Denner}},
  \bibinfo {author} {\bibfnamefont {Nana}\ \bibnamefont {Shumiya}}, \bibinfo
  {author} {\bibfnamefont {Brenden~R}\ \bibnamefont {Ortiz}}, \bibinfo {author}
  {\bibfnamefont {Gang}\ \bibnamefont {Xu}}, \bibinfo {author} {\bibfnamefont
  {Zurab}\ \bibnamefont {Guguchia}}, \bibinfo {author} {\bibfnamefont {Junyi}\
  \bibnamefont {He}}, \bibinfo {author} {\bibfnamefont {Md~Shafayat}\
  \bibnamefont {Hossain}}, \bibinfo {author} {\bibfnamefont {Xiaoxiong}\
  \bibnamefont {Liu}},  \emph {et~al.},\ }\bibfield  {title} {\enquote
  {\bibinfo {title} {{Unconventional chiral charge order in kagome
  superconductor KV$_3$Sb$_5$}},}\ }\href {\doibase 10.1038/s41563-021-01034-y}
  {\bibfield  {journal} {\bibinfo  {journal} {Nature Materials}\ }\textbf
  {\bibinfo {volume} {20}},\ \bibinfo {pages} {1353--1357} (\bibinfo {year}
  {2021})}\BibitemShut {NoStop}%
\bibitem [{\citenamefont {Tan}\ \emph {et~al.}(2021)\citenamefont {Tan},
  \citenamefont {Liu}, \citenamefont {Wang},\ and\ \citenamefont
  {Yan}}]{tan2021charge}%
  \BibitemOpen
  \bibfield  {author} {\bibinfo {author} {\bibfnamefont {Hengxin}\ \bibnamefont
  {Tan}}, \bibinfo {author} {\bibfnamefont {Yizhou}\ \bibnamefont {Liu}},
  \bibinfo {author} {\bibfnamefont {Ziqiang}\ \bibnamefont {Wang}}, \ and\
  \bibinfo {author} {\bibfnamefont {Binghai}\ \bibnamefont {Yan}},\ }\bibfield
  {title} {\enquote {\bibinfo {title} {Charge density waves and electronic
  properties of superconducting kagome metals},}\ }\href {\doibase
  10.1103/PhysRevLett.127.046401} {\bibfield  {journal} {\bibinfo  {journal}
  {Phys. Rev. Lett.}\ }\textbf {\bibinfo {volume} {127}},\ \bibinfo {pages}
  {046401} (\bibinfo {year} {2021})}\BibitemShut {NoStop}%
\bibitem [{\citenamefont {Liang}\ \emph {et~al.}(2021)\citenamefont {Liang},
  \citenamefont {Hou}, \citenamefont {Zhang}, \citenamefont {Ma}, \citenamefont
  {Wu}, \citenamefont {Zhang}, \citenamefont {Yu}, \citenamefont {Ying},
  \citenamefont {Jiang}, \citenamefont {Shan}, \citenamefont {Wang},\ and\
  \citenamefont {Chen}}]{Liang2021three}%
  \BibitemOpen
  \bibfield  {author} {\bibinfo {author} {\bibfnamefont {Zuowei}\ \bibnamefont
  {Liang}}, \bibinfo {author} {\bibfnamefont {Xingyuan}\ \bibnamefont {Hou}},
  \bibinfo {author} {\bibfnamefont {Fan}\ \bibnamefont {Zhang}}, \bibinfo
  {author} {\bibfnamefont {Wanru}\ \bibnamefont {Ma}}, \bibinfo {author}
  {\bibfnamefont {Ping}\ \bibnamefont {Wu}}, \bibinfo {author} {\bibfnamefont
  {Zongyuan}\ \bibnamefont {Zhang}}, \bibinfo {author} {\bibfnamefont
  {Fanghang}\ \bibnamefont {Yu}}, \bibinfo {author} {\bibfnamefont {J.-J.}\
  \bibnamefont {Ying}}, \bibinfo {author} {\bibfnamefont {Kun}\ \bibnamefont
  {Jiang}}, \bibinfo {author} {\bibfnamefont {Lei}\ \bibnamefont {Shan}},
  \bibinfo {author} {\bibfnamefont {Zhenyu}\ \bibnamefont {Wang}}, \ and\
  \bibinfo {author} {\bibfnamefont {X.-H.}\ \bibnamefont {Chen}},\ }\bibfield
  {title} {\enquote {\bibinfo {title} {{Three-Dimensional Charge Density Wave
  and Surface-Dependent Vortex-Core States in a Kagome Superconductor
  ${\mathrm{CsV}}_{3}{\mathrm{Sb}}_{5}$}},}\ }\href {\doibase
  10.1103/PhysRevX.11.031026} {\bibfield  {journal} {\bibinfo  {journal} {Phys.
  Rev. X}\ }\textbf {\bibinfo {volume} {11}},\ \bibinfo {pages} {031026}
  (\bibinfo {year} {2021})}\BibitemShut {NoStop}%
\bibitem [{\citenamefont {Chen}\ \emph {et~al.}(2021)\citenamefont {Chen},
  \citenamefont {Yang}, \citenamefont {Hu}, \citenamefont {Zhao}, \citenamefont
  {Yuan}, \citenamefont {Xing}, \citenamefont {Qian}, \citenamefont {Huang},
  \citenamefont {Li}, \citenamefont {Ye} \emph {et~al.}}]{chen2021roton}%
  \BibitemOpen
  \bibfield  {author} {\bibinfo {author} {\bibfnamefont {Hui}\ \bibnamefont
  {Chen}}, \bibinfo {author} {\bibfnamefont {Haitao}\ \bibnamefont {Yang}},
  \bibinfo {author} {\bibfnamefont {Bin}\ \bibnamefont {Hu}}, \bibinfo {author}
  {\bibfnamefont {Zhen}\ \bibnamefont {Zhao}}, \bibinfo {author} {\bibfnamefont
  {Jie}\ \bibnamefont {Yuan}}, \bibinfo {author} {\bibfnamefont {Yuqing}\
  \bibnamefont {Xing}}, \bibinfo {author} {\bibfnamefont {Guojian}\
  \bibnamefont {Qian}}, \bibinfo {author} {\bibfnamefont {Zihao}\ \bibnamefont
  {Huang}}, \bibinfo {author} {\bibfnamefont {Geng}\ \bibnamefont {Li}},
  \bibinfo {author} {\bibfnamefont {Yuhan}\ \bibnamefont {Ye}},  \emph
  {et~al.},\ }\bibfield  {title} {\enquote {\bibinfo {title} {Roton pair
  density wave in a strong-coupling kagome superconductor},}\ }\href {\doibase
  10.1038/s41586-021-03983-5} {\bibfield  {journal} {\bibinfo  {journal}
  {Nature}\ }\textbf {\bibinfo {volume} {599}},\ \bibinfo {pages} {222--228}
  (\bibinfo {year} {2021})}\BibitemShut {NoStop}%
\bibitem [{\citenamefont {Teng}\ \emph {et~al.}(2022)\citenamefont {Teng},
  \citenamefont {Chen}, \citenamefont {Ye}, \citenamefont {Rosenberg},
  \citenamefont {Liu}, \citenamefont {Yin}, \citenamefont {Jiang},
  \citenamefont {Oh}, \citenamefont {Hasan}, \citenamefont {Neubauer} \emph
  {et~al.}}]{teng2022discovery}%
  \BibitemOpen
  \bibfield  {author} {\bibinfo {author} {\bibfnamefont {Xiaokun}\ \bibnamefont
  {Teng}}, \bibinfo {author} {\bibfnamefont {Lebing}\ \bibnamefont {Chen}},
  \bibinfo {author} {\bibfnamefont {Feng}\ \bibnamefont {Ye}}, \bibinfo
  {author} {\bibfnamefont {Elliott}\ \bibnamefont {Rosenberg}}, \bibinfo
  {author} {\bibfnamefont {Zhaoyu}\ \bibnamefont {Liu}}, \bibinfo {author}
  {\bibfnamefont {Jia-Xin}\ \bibnamefont {Yin}}, \bibinfo {author}
  {\bibfnamefont {Yu-Xiao}\ \bibnamefont {Jiang}}, \bibinfo {author}
  {\bibfnamefont {Ji~Seop}\ \bibnamefont {Oh}}, \bibinfo {author}
  {\bibfnamefont {M~Zahid}\ \bibnamefont {Hasan}}, \bibinfo {author}
  {\bibfnamefont {Kelly~J}\ \bibnamefont {Neubauer}},  \emph {et~al.},\
  }\bibfield  {title} {\enquote {\bibinfo {title} {Discovery of charge density
  wave in a kagome lattice antiferromagnet},}\ }\href {\doibase
  10.1038/s41586-022-05034-z} {\bibfield  {journal} {\bibinfo  {journal}
  {Nature}\ }\textbf {\bibinfo {volume} {609}},\ \bibinfo {pages} {490--495}
  (\bibinfo {year} {2022})}\BibitemShut {NoStop}%
\bibitem [{\citenamefont {Teng}\ \emph {et~al.}(2023)\citenamefont {Teng},
  \citenamefont {Oh}, \citenamefont {Tan}, \citenamefont {Chen}, \citenamefont
  {Huang}, \citenamefont {Gao}, \citenamefont {Yin}, \citenamefont {Chu},
  \citenamefont {Hashimoto}, \citenamefont {Lu} \emph
  {et~al.}}]{teng2023magnetism}%
  \BibitemOpen
  \bibfield  {author} {\bibinfo {author} {\bibfnamefont {Xiaokun}\ \bibnamefont
  {Teng}}, \bibinfo {author} {\bibfnamefont {Ji~Seop}\ \bibnamefont {Oh}},
  \bibinfo {author} {\bibfnamefont {Hengxin}\ \bibnamefont {Tan}}, \bibinfo
  {author} {\bibfnamefont {Lebing}\ \bibnamefont {Chen}}, \bibinfo {author}
  {\bibfnamefont {Jianwei}\ \bibnamefont {Huang}}, \bibinfo {author}
  {\bibfnamefont {Bin}\ \bibnamefont {Gao}}, \bibinfo {author} {\bibfnamefont
  {Jia-Xin}\ \bibnamefont {Yin}}, \bibinfo {author} {\bibfnamefont {Jiun-Haw}\
  \bibnamefont {Chu}}, \bibinfo {author} {\bibfnamefont {Makoto}\ \bibnamefont
  {Hashimoto}}, \bibinfo {author} {\bibfnamefont {Donghui}\ \bibnamefont {Lu}},
   \emph {et~al.},\ }\bibfield  {title} {\enquote {\bibinfo {title} {Magnetism
  and charge density wave order in kagome fege},}\ }\href {\doibase
  10.1038/s41567-023-01985-w} {\bibfield  {journal} {\bibinfo  {journal}
  {Nature Physics}\ }\textbf {\bibinfo {volume} {19}},\ \bibinfo {pages}
  {814--822} (\bibinfo {year} {2023})}\BibitemShut {NoStop}%
\bibitem [{\citenamefont {Miao}\ \emph {et~al.}(2023)\citenamefont {Miao},
  \citenamefont {Zhang}, \citenamefont {Li}, \citenamefont {Fabbris},
  \citenamefont {Said}, \citenamefont {Tartaglia}, \citenamefont {Yilmaz},
  \citenamefont {Vescovo}, \citenamefont {Yin}, \citenamefont {Murakami} \emph
  {et~al.}}]{miao2023signature}%
  \BibitemOpen
  \bibfield  {author} {\bibinfo {author} {\bibfnamefont {H}~\bibnamefont
  {Miao}}, \bibinfo {author} {\bibfnamefont {TT}~\bibnamefont {Zhang}},
  \bibinfo {author} {\bibfnamefont {HX}~\bibnamefont {Li}}, \bibinfo {author}
  {\bibfnamefont {G}~\bibnamefont {Fabbris}}, \bibinfo {author} {\bibfnamefont
  {AH}~\bibnamefont {Said}}, \bibinfo {author} {\bibfnamefont {R}~\bibnamefont
  {Tartaglia}}, \bibinfo {author} {\bibfnamefont {T}~\bibnamefont {Yilmaz}},
  \bibinfo {author} {\bibfnamefont {E}~\bibnamefont {Vescovo}}, \bibinfo
  {author} {\bibfnamefont {J-X}\ \bibnamefont {Yin}}, \bibinfo {author}
  {\bibfnamefont {S}~\bibnamefont {Murakami}},  \emph {et~al.},\ }\bibfield
  {title} {\enquote {\bibinfo {title} {Signature of spin-phonon coupling driven
  charge density wave in a kagome magnet},}\ }\href {\doibase
  10.1038/s41467-023-41957-5} {\bibfield  {journal} {\bibinfo  {journal}
  {Nature Communications}\ }\textbf {\bibinfo {volume} {14}},\ \bibinfo {pages}
  {6183} (\bibinfo {year} {2023})}\BibitemShut {NoStop}%
\bibitem [{\citenamefont {Wang}(2023)}]{wang2023enhanced}%
  \BibitemOpen
  \bibfield  {author} {\bibinfo {author} {\bibfnamefont {Yilin}\ \bibnamefont
  {Wang}},\ }\bibfield  {title} {\enquote {\bibinfo {title} {{Enhanced
  spin-polarization via partial Ge-dimerization as the driving force of the
  charge density wave in FeGe}},}\ }\href {\doibase
  10.1103/PhysRevMaterials.7.104006} {\bibfield  {journal} {\bibinfo  {journal}
  {Phys. Rev. Mater.}\ }\textbf {\bibinfo {volume} {7}},\ \bibinfo {pages}
  {104006} (\bibinfo {year} {2023})}\BibitemShut {NoStop}%
\bibitem [{\citenamefont {Chen}\ \emph {et~al.}(2023)\citenamefont {Chen},
  \citenamefont {Wu}, \citenamefont {Zhou}, \citenamefont {Zhang},
  \citenamefont {Yin}, \citenamefont {Li}, \citenamefont {Li}, \citenamefont
  {Gong}, \citenamefont {He}, \citenamefont {Chai}, \citenamefont {Zhou},
  \citenamefont {Wang}, \citenamefont {Wang}, \citenamefont {Yan},\ and\
  \citenamefont {Feng}}]{chen2023longranged}%
  \BibitemOpen
  \bibfield  {author} {\bibinfo {author} {\bibfnamefont {Ziyuan}\ \bibnamefont
  {Chen}}, \bibinfo {author} {\bibfnamefont {Xueliang}\ \bibnamefont {Wu}},
  \bibinfo {author} {\bibfnamefont {Shiming}\ \bibnamefont {Zhou}}, \bibinfo
  {author} {\bibfnamefont {Jiakang}\ \bibnamefont {Zhang}}, \bibinfo {author}
  {\bibfnamefont {Ruotong}\ \bibnamefont {Yin}}, \bibinfo {author}
  {\bibfnamefont {Yuanji}\ \bibnamefont {Li}}, \bibinfo {author} {\bibfnamefont
  {Mingzhe}\ \bibnamefont {Li}}, \bibinfo {author} {\bibfnamefont {Jiashuo}\
  \bibnamefont {Gong}}, \bibinfo {author} {\bibfnamefont {Mingquan}\
  \bibnamefont {He}}, \bibinfo {author} {\bibfnamefont {Yisheng}\ \bibnamefont
  {Chai}}, \bibinfo {author} {\bibfnamefont {Xiaoyuan}\ \bibnamefont {Zhou}},
  \bibinfo {author} {\bibfnamefont {Yilin}\ \bibnamefont {Wang}}, \bibinfo
  {author} {\bibfnamefont {Aifeng}\ \bibnamefont {Wang}}, \bibinfo {author}
  {\bibfnamefont {Ya-Jun}\ \bibnamefont {Yan}}, \ and\ \bibinfo {author}
  {\bibfnamefont {Dong-Lai}\ \bibnamefont {Feng}},\ }\bibfield  {title}
  {\enquote {\bibinfo {title} {Long-ranged charge order conspired by magnetism
  and lattice in an antiferromagnetic kagome metal},}\ }\href
  {https://arxiv.org/abs/2307.07990} {\bibfield  {journal} {\bibinfo  {journal}
  {arXiv:2307.07990}\ } (\bibinfo {year} {2023})}\BibitemShut {NoStop}%
\bibitem [{\citenamefont {Ghimire}\ \emph {et~al.}(2020)\citenamefont
  {Ghimire}, \citenamefont {Dally}, \citenamefont {Poudel}, \citenamefont
  {Jones}, \citenamefont {Michel}, \citenamefont {Magar}, \citenamefont
  {Bleuel}, \citenamefont {McGuire}, \citenamefont {Jiang}, \citenamefont
  {Mitchell}, \citenamefont {Lynn},\ and\ \citenamefont
  {Mazin}}]{Ghimire2020Competing}%
  \BibitemOpen
  \bibfield  {author} {\bibinfo {author} {\bibfnamefont {Nirmal~J.}\
  \bibnamefont {Ghimire}}, \bibinfo {author} {\bibfnamefont {Rebecca~L.}\
  \bibnamefont {Dally}}, \bibinfo {author} {\bibfnamefont {L.}~\bibnamefont
  {Poudel}}, \bibinfo {author} {\bibfnamefont {D.~C.}\ \bibnamefont {Jones}},
  \bibinfo {author} {\bibfnamefont {D.}~\bibnamefont {Michel}}, \bibinfo
  {author} {\bibfnamefont {N.~Thapa}\ \bibnamefont {Magar}}, \bibinfo {author}
  {\bibfnamefont {M.}~\bibnamefont {Bleuel}}, \bibinfo {author} {\bibfnamefont
  {Michael~A.}\ \bibnamefont {McGuire}}, \bibinfo {author} {\bibfnamefont
  {J.~S.}\ \bibnamefont {Jiang}}, \bibinfo {author} {\bibfnamefont {J.~F.}\
  \bibnamefont {Mitchell}}, \bibinfo {author} {\bibfnamefont {Jeffrey~W.}\
  \bibnamefont {Lynn}}, \ and\ \bibinfo {author} {\bibfnamefont {I.~I.}\
  \bibnamefont {Mazin}},\ }\bibfield  {title} {\enquote {\bibinfo {title}
  {{Competing magnetic phases and fluctuation-driven scalar spin chirality in
  the kagome metal YMn$_6$Sn$_6$}},}\ }\href {\doibase 10.1126/sciadv.abe2680}
  {\bibfield  {journal} {\bibinfo  {journal} {Science Advances}\ }\textbf
  {\bibinfo {volume} {6}},\ \bibinfo {pages} {eabe2680} (\bibinfo {year}
  {2020})}\BibitemShut {NoStop}%
\bibitem [{\citenamefont {Riberolles}\ \emph {et~al.}(2022)\citenamefont
  {Riberolles}, \citenamefont {Slade}, \citenamefont {Abernathy}, \citenamefont
  {Granroth}, \citenamefont {Li}, \citenamefont {Lee}, \citenamefont
  {Canfield}, \citenamefont {Ueland}, \citenamefont {Ke},\ and\ \citenamefont
  {McQueeney}}]{Riberolles2022low}%
  \BibitemOpen
  \bibfield  {author} {\bibinfo {author} {\bibfnamefont {S.~X.~M.}\
  \bibnamefont {Riberolles}}, \bibinfo {author} {\bibfnamefont {Tyler~J.}\
  \bibnamefont {Slade}}, \bibinfo {author} {\bibfnamefont {D.~L.}\ \bibnamefont
  {Abernathy}}, \bibinfo {author} {\bibfnamefont {G.~E.}\ \bibnamefont
  {Granroth}}, \bibinfo {author} {\bibfnamefont {Bing}\ \bibnamefont {Li}},
  \bibinfo {author} {\bibfnamefont {Y.}~\bibnamefont {Lee}}, \bibinfo {author}
  {\bibfnamefont {P.~C.}\ \bibnamefont {Canfield}}, \bibinfo {author}
  {\bibfnamefont {B.~G.}\ \bibnamefont {Ueland}}, \bibinfo {author}
  {\bibfnamefont {Liqin}\ \bibnamefont {Ke}}, \ and\ \bibinfo {author}
  {\bibfnamefont {R.~J.}\ \bibnamefont {McQueeney}},\ }\bibfield  {title}
  {\enquote {\bibinfo {title} {{Low-Temperature Competing Magnetic Energy
  Scales in the Topological Ferrimagnet
  ${\mathrm{TbMn}}_{6}{\mathrm{Sn}}_{6}$}},}\ }\href {\doibase
  10.1103/PhysRevX.12.021043} {\bibfield  {journal} {\bibinfo  {journal} {Phys.
  Rev. X}\ }\textbf {\bibinfo {volume} {12}},\ \bibinfo {pages} {021043}
  (\bibinfo {year} {2022})}\BibitemShut {NoStop}%
\bibitem [{\citenamefont {Yin}\ \emph {et~al.}(2020)\citenamefont {Yin},
  \citenamefont {Ma}, \citenamefont {Cochran}, \citenamefont {Xu},
  \citenamefont {Zhang}, \citenamefont {Tien}, \citenamefont {Shumiya},
  \citenamefont {Cheng}, \citenamefont {Jiang}, \citenamefont {Lian},
  \citenamefont {Song}, \citenamefont {Chang}, \citenamefont {Belopolski},
  \citenamefont {Multer}, \citenamefont {Litskevich}, \citenamefont {Cheng},
  \citenamefont {Yang}, \citenamefont {Swidler}, \citenamefont {Zhou},
  \citenamefont {Lin}, \citenamefont {Neupert}, \citenamefont {Wang},
  \citenamefont {Yao}, \citenamefont {Chang}, \citenamefont {Jia},\ and\
  \citenamefont {Hasan}}]{yin2020quantum}%
  \BibitemOpen
  \bibfield  {author} {\bibinfo {author} {\bibfnamefont {Jia-Xin}\ \bibnamefont
  {Yin}}, \bibinfo {author} {\bibfnamefont {Wenlong}\ \bibnamefont {Ma}},
  \bibinfo {author} {\bibfnamefont {Tyler~A}\ \bibnamefont {Cochran}}, \bibinfo
  {author} {\bibfnamefont {Xitong}\ \bibnamefont {Xu}}, \bibinfo {author}
  {\bibfnamefont {Songtian~S}\ \bibnamefont {Zhang}}, \bibinfo {author}
  {\bibfnamefont {Hung-Ju}\ \bibnamefont {Tien}}, \bibinfo {author}
  {\bibfnamefont {Nana}\ \bibnamefont {Shumiya}}, \bibinfo {author}
  {\bibfnamefont {Guangming}\ \bibnamefont {Cheng}}, \bibinfo {author}
  {\bibfnamefont {Kun}\ \bibnamefont {Jiang}}, \bibinfo {author} {\bibfnamefont
  {Biao}\ \bibnamefont {Lian}}, \bibinfo {author} {\bibfnamefont {Zhida}\
  \bibnamefont {Song}}, \bibinfo {author} {\bibfnamefont {Guoqing}\
  \bibnamefont {Chang}}, \bibinfo {author} {\bibfnamefont {Ilya}\ \bibnamefont
  {Belopolski}}, \bibinfo {author} {\bibfnamefont {Daniel}\ \bibnamefont
  {Multer}}, \bibinfo {author} {\bibfnamefont {Maksim}\ \bibnamefont
  {Litskevich}}, \bibinfo {author} {\bibfnamefont {Zi-Jia}\ \bibnamefont
  {Cheng}}, \bibinfo {author} {\bibfnamefont {Xian~P.}\ \bibnamefont {Yang}},
  \bibinfo {author} {\bibfnamefont {Bianca}\ \bibnamefont {Swidler}}, \bibinfo
  {author} {\bibfnamefont {Huibin}\ \bibnamefont {Zhou}}, \bibinfo {author}
  {\bibfnamefont {Hsin}\ \bibnamefont {Lin}}, \bibinfo {author} {\bibfnamefont
  {Titus}\ \bibnamefont {Neupert}}, \bibinfo {author} {\bibfnamefont {Ziqiang}\
  \bibnamefont {Wang}}, \bibinfo {author} {\bibfnamefont {Nan}\ \bibnamefont
  {Yao}}, \bibinfo {author} {\bibfnamefont {Tay-Rong}\ \bibnamefont {Chang}},
  \bibinfo {author} {\bibfnamefont {Shuang}\ \bibnamefont {Jia}}, \ and\
  \bibinfo {author} {\bibfnamefont {M.~Zahid}\ \bibnamefont {Hasan}},\
  }\bibfield  {title} {\enquote {\bibinfo {title} {{Quantum-limit Chern
  topological magnetism in TbMn$_6$Sn$_6$}},}\ }\href {\doibase
  10.1038/s41586-018-0502-7} {\bibfield  {journal} {\bibinfo  {journal}
  {Nature}\ }\textbf {\bibinfo {volume} {583}},\ \bibinfo {pages} {533--536}
  (\bibinfo {year} {2020})}\BibitemShut {NoStop}%
\bibitem [{\citenamefont {Zhang}\ \emph {et~al.}(2020)\citenamefont {Zhang},
  \citenamefont {Feng}, \citenamefont {Heitmann}, \citenamefont {Kolesnikov},
  \citenamefont {Stone}, \citenamefont {Lu},\ and\ \citenamefont
  {Ke}}]{zhang2020Topological}%
  \BibitemOpen
  \bibfield  {author} {\bibinfo {author} {\bibfnamefont {H.}~\bibnamefont
  {Zhang}}, \bibinfo {author} {\bibfnamefont {X.}~\bibnamefont {Feng}},
  \bibinfo {author} {\bibfnamefont {T.}~\bibnamefont {Heitmann}}, \bibinfo
  {author} {\bibfnamefont {A.~I.}\ \bibnamefont {Kolesnikov}}, \bibinfo
  {author} {\bibfnamefont {M.~B.}\ \bibnamefont {Stone}}, \bibinfo {author}
  {\bibfnamefont {Y.-M.}\ \bibnamefont {Lu}}, \ and\ \bibinfo {author}
  {\bibfnamefont {X.}~\bibnamefont {Ke}},\ }\bibfield  {title} {\enquote
  {\bibinfo {title} {Topological magnon bands in a room-temperature kagome
  magnet},}\ }\href {\doibase 10.1103/PhysRevB.101.100405} {\bibfield
  {journal} {\bibinfo  {journal} {Phys. Rev. B}\ }\textbf {\bibinfo {volume}
  {101}},\ \bibinfo {pages} {100405} (\bibinfo {year} {2020})}\BibitemShut
  {NoStop}%
\bibitem [{\citenamefont {Li}\ \emph {et~al.}(2021)\citenamefont {Li},
  \citenamefont {Wang}, \citenamefont {Wang}, \citenamefont {Yuan},
  \citenamefont {Song}, \citenamefont {Lou}, \citenamefont {Liu}, \citenamefont
  {Huang}, \citenamefont {Liu}, \citenamefont {Lei} \emph
  {et~al.}}]{li2021dirac}%
  \BibitemOpen
  \bibfield  {author} {\bibinfo {author} {\bibfnamefont {Man}\ \bibnamefont
  {Li}}, \bibinfo {author} {\bibfnamefont {Qi}~\bibnamefont {Wang}}, \bibinfo
  {author} {\bibfnamefont {Guangwei}\ \bibnamefont {Wang}}, \bibinfo {author}
  {\bibfnamefont {Zhihong}\ \bibnamefont {Yuan}}, \bibinfo {author}
  {\bibfnamefont {Wenhua}\ \bibnamefont {Song}}, \bibinfo {author}
  {\bibfnamefont {Rui}\ \bibnamefont {Lou}}, \bibinfo {author} {\bibfnamefont
  {Zhengtai}\ \bibnamefont {Liu}}, \bibinfo {author} {\bibfnamefont {Yaobo}\
  \bibnamefont {Huang}}, \bibinfo {author} {\bibfnamefont {Zhonghao}\
  \bibnamefont {Liu}}, \bibinfo {author} {\bibfnamefont {Hechang}\ \bibnamefont
  {Lei}},  \emph {et~al.},\ }\bibfield  {title} {\enquote {\bibinfo {title}
  {{Dirac cone, flat band and saddle point in kagome magnet YMn$_6$Sn$_6$}},}\
  }\href {\doibase 10.1038/s41467-021-23536-8} {\bibfield  {journal} {\bibinfo
  {journal} {Nature Communications}\ }\textbf {\bibinfo {volume} {12}},\
  \bibinfo {pages} {3129} (\bibinfo {year} {2021})}\BibitemShut {NoStop}%
\bibitem [{\citenamefont {Xu}\ \emph {et~al.}(2022)\citenamefont {Xu},
  \citenamefont {Yin}, \citenamefont {Ma}, \citenamefont {Tien}, \citenamefont
  {Qiang}, \citenamefont {Reddy}, \citenamefont {Zhou}, \citenamefont {Shen},
  \citenamefont {Lu}, \citenamefont {Chang} \emph
  {et~al.}}]{xu2022topological}%
  \BibitemOpen
  \bibfield  {author} {\bibinfo {author} {\bibfnamefont {Xitong}\ \bibnamefont
  {Xu}}, \bibinfo {author} {\bibfnamefont {Jia-Xin}\ \bibnamefont {Yin}},
  \bibinfo {author} {\bibfnamefont {Wenlong}\ \bibnamefont {Ma}}, \bibinfo
  {author} {\bibfnamefont {Hung-Ju}\ \bibnamefont {Tien}}, \bibinfo {author}
  {\bibfnamefont {Xiao-Bin}\ \bibnamefont {Qiang}}, \bibinfo {author}
  {\bibfnamefont {PV}~\bibnamefont {Reddy}}, \bibinfo {author} {\bibfnamefont
  {Huibin}\ \bibnamefont {Zhou}}, \bibinfo {author} {\bibfnamefont {Jie}\
  \bibnamefont {Shen}}, \bibinfo {author} {\bibfnamefont {Hai-Zhou}\
  \bibnamefont {Lu}}, \bibinfo {author} {\bibfnamefont {Tay-Rong}\ \bibnamefont
  {Chang}},  \emph {et~al.},\ }\bibfield  {title} {\enquote {\bibinfo {title}
  {{Topological charge-entropy scaling in kagome Chern magnet
  TbMn$_6$Sn$_6$}},}\ }\href {\doibase 10.1038/s41467-022-28796-6} {\bibfield
  {journal} {\bibinfo  {journal} {Nature Communications}\ }\textbf {\bibinfo
  {volume} {13}},\ \bibinfo {pages} {1197} (\bibinfo {year}
  {2022})}\BibitemShut {NoStop}%
\bibitem [{\citenamefont {Li}\ \emph {et~al.}(2022)\citenamefont {Li},
  \citenamefont {Zhao}, \citenamefont {Jiang}, \citenamefont {Wang},
  \citenamefont {Yin}, \citenamefont {Zhao}, \citenamefont {Liu}, \citenamefont
  {Wang}, \citenamefont {Lei},\ and\ \citenamefont
  {Zeljkovic}}]{li2022manipulation}%
  \BibitemOpen
  \bibfield  {author} {\bibinfo {author} {\bibfnamefont {Hong}\ \bibnamefont
  {Li}}, \bibinfo {author} {\bibfnamefont {He}~\bibnamefont {Zhao}}, \bibinfo
  {author} {\bibfnamefont {Kun}\ \bibnamefont {Jiang}}, \bibinfo {author}
  {\bibfnamefont {Qi}~\bibnamefont {Wang}}, \bibinfo {author} {\bibfnamefont
  {Qiangwei}\ \bibnamefont {Yin}}, \bibinfo {author} {\bibfnamefont
  {Ning-Ning}\ \bibnamefont {Zhao}}, \bibinfo {author} {\bibfnamefont {Kai}\
  \bibnamefont {Liu}}, \bibinfo {author} {\bibfnamefont {Ziqiang}\ \bibnamefont
  {Wang}}, \bibinfo {author} {\bibfnamefont {Hechang}\ \bibnamefont {Lei}}, \
  and\ \bibinfo {author} {\bibfnamefont {Ilija}\ \bibnamefont {Zeljkovic}},\
  }\bibfield  {title} {\enquote {\bibinfo {title} {{Manipulation of Dirac band
  curvature and momentum-dependent g factor in a kagome magnet}},}\ }\href
  {\doibase 10.1038/s41567-022-01558-3} {\bibfield  {journal} {\bibinfo
  {journal} {Nature Physics}\ }\textbf {\bibinfo {volume} {18}},\ \bibinfo
  {pages} {644--649} (\bibinfo {year} {2022})}\BibitemShut {NoStop}%
\bibitem [{\citenamefont {Asaba}\ \emph {et~al.}(2020)\citenamefont {Asaba},
  \citenamefont {Thomas}, \citenamefont {Curtis}, \citenamefont {Thompson},
  \citenamefont {Bauer},\ and\ \citenamefont {Ronning}}]{Asaba2020anomalous}%
  \BibitemOpen
  \bibfield  {author} {\bibinfo {author} {\bibfnamefont {T.}~\bibnamefont
  {Asaba}}, \bibinfo {author} {\bibfnamefont {S.~M.}\ \bibnamefont {Thomas}},
  \bibinfo {author} {\bibfnamefont {M.}~\bibnamefont {Curtis}}, \bibinfo
  {author} {\bibfnamefont {J.~D.}\ \bibnamefont {Thompson}}, \bibinfo {author}
  {\bibfnamefont {E.~D.}\ \bibnamefont {Bauer}}, \ and\ \bibinfo {author}
  {\bibfnamefont {F.}~\bibnamefont {Ronning}},\ }\bibfield  {title} {\enquote
  {\bibinfo {title} {{Anomalous Hall effect in the kagome ferrimagnet
  ${\mathrm{GdMn}}_{6}{\mathrm{Sn}}_{6}$}},}\ }\href {\doibase
  10.1103/PhysRevB.101.174415} {\bibfield  {journal} {\bibinfo  {journal}
  {Phys. Rev. B}\ }\textbf {\bibinfo {volume} {101}},\ \bibinfo {pages}
  {174415} (\bibinfo {year} {2020})}\BibitemShut {NoStop}%
\bibitem [{\citenamefont {Ma}\ \emph {et~al.}(2021)\citenamefont {Ma},
  \citenamefont {Xu}, \citenamefont {Yin}, \citenamefont {Yang}, \citenamefont
  {Zhou}, \citenamefont {Cheng}, \citenamefont {Huang}, \citenamefont {Qu},
  \citenamefont {Wang}, \citenamefont {Hasan},\ and\ \citenamefont
  {Jia}}]{ma2021rare}%
  \BibitemOpen
  \bibfield  {author} {\bibinfo {author} {\bibfnamefont {Wenlong}\ \bibnamefont
  {Ma}}, \bibinfo {author} {\bibfnamefont {Xitong}\ \bibnamefont {Xu}},
  \bibinfo {author} {\bibfnamefont {Jia-Xin}\ \bibnamefont {Yin}}, \bibinfo
  {author} {\bibfnamefont {Hui}\ \bibnamefont {Yang}}, \bibinfo {author}
  {\bibfnamefont {Huibin}\ \bibnamefont {Zhou}}, \bibinfo {author}
  {\bibfnamefont {Zi-Jia}\ \bibnamefont {Cheng}}, \bibinfo {author}
  {\bibfnamefont {Yuqing}\ \bibnamefont {Huang}}, \bibinfo {author}
  {\bibfnamefont {Zhe}\ \bibnamefont {Qu}}, \bibinfo {author} {\bibfnamefont
  {Fa}~\bibnamefont {Wang}}, \bibinfo {author} {\bibfnamefont {M.~Zahid}\
  \bibnamefont {Hasan}}, \ and\ \bibinfo {author} {\bibfnamefont {Shuang}\
  \bibnamefont {Jia}},\ }\bibfield  {title} {\enquote {\bibinfo {title} {{Rare
  Earth Engineering in $R{\mathrm{Mn}}_{6}{\mathrm{Sn}}_{6}$
  ($R=\text{Gd}\text{\ensuremath{-}}\text{Tm}$, Lu) Topological Kagome
  Magnets}},}\ }\href {\doibase 10.1103/PhysRevLett.126.246602} {\bibfield
  {journal} {\bibinfo  {journal} {Phys. Rev. Lett.}\ }\textbf {\bibinfo
  {volume} {126}},\ \bibinfo {pages} {246602} (\bibinfo {year}
  {2021})}\BibitemShut {NoStop}%
\bibitem [{\citenamefont {Gao}\ \emph {et~al.}(2021)\citenamefont {Gao},
  \citenamefont {Shen}, \citenamefont {Wang}, \citenamefont {Shi},
  \citenamefont {Zhao}, \citenamefont {Li}, \citenamefont {Cao}, \citenamefont
  {Pei}, \citenamefont {Ge}, \citenamefont {Li}, \citenamefont {Li},
  \citenamefont {Chen}, \citenamefont {Yan},\ and\ \citenamefont
  {Qi}}]{Gao2021anomalous}%
  \BibitemOpen
  \bibfield  {author} {\bibinfo {author} {\bibfnamefont {Lingling}\
  \bibnamefont {Gao}}, \bibinfo {author} {\bibfnamefont {Shiwei}\ \bibnamefont
  {Shen}}, \bibinfo {author} {\bibfnamefont {Qi}~\bibnamefont {Wang}}, \bibinfo
  {author} {\bibfnamefont {Wujun}\ \bibnamefont {Shi}}, \bibinfo {author}
  {\bibfnamefont {Yi}~\bibnamefont {Zhao}}, \bibinfo {author} {\bibfnamefont
  {Changhua}\ \bibnamefont {Li}}, \bibinfo {author} {\bibfnamefont {Weizheng}\
  \bibnamefont {Cao}}, \bibinfo {author} {\bibfnamefont {Cuiying}\ \bibnamefont
  {Pei}}, \bibinfo {author} {\bibfnamefont {Jun-Yi}\ \bibnamefont {Ge}},
  \bibinfo {author} {\bibfnamefont {Gang}\ \bibnamefont {Li}}, \bibinfo
  {author} {\bibfnamefont {Jun}\ \bibnamefont {Li}}, \bibinfo {author}
  {\bibfnamefont {Yulin}\ \bibnamefont {Chen}}, \bibinfo {author}
  {\bibfnamefont {Shichao}\ \bibnamefont {Yan}}, \ and\ \bibinfo {author}
  {\bibfnamefont {Yanpeng}\ \bibnamefont {Qi}},\ }\bibfield  {title} {\enquote
  {\bibinfo {title} {{Anomalous Hall effect in ferrimagnetic metal
  RMn$_6$Sn$_6$ (R = Tb, Dy, Ho) with clean Mn kagome lattice}},}\ }\href
  {\doibase 10.1063/5.0061260} {\bibfield  {journal} {\bibinfo  {journal}
  {Applied Physics Letters}\ }\textbf {\bibinfo {volume} {119}},\ \bibinfo
  {pages} {092405} (\bibinfo {year} {2021})}\BibitemShut {NoStop}%
\bibitem [{\citenamefont {Zhang}\ \emph
  {et~al.}(2022{\natexlab{a}})\citenamefont {Zhang}, \citenamefont {Koo},
  \citenamefont {Xu}, \citenamefont {Sretenovic}, \citenamefont {Yan},\ and\
  \citenamefont {Ke}}]{zhang2022exchange}%
  \BibitemOpen
  \bibfield  {author} {\bibinfo {author} {\bibfnamefont {Heda}\ \bibnamefont
  {Zhang}}, \bibinfo {author} {\bibfnamefont {Jahyun}\ \bibnamefont {Koo}},
  \bibinfo {author} {\bibfnamefont {Chunqiang}\ \bibnamefont {Xu}}, \bibinfo
  {author} {\bibfnamefont {Milos}\ \bibnamefont {Sretenovic}}, \bibinfo
  {author} {\bibfnamefont {Binghai}\ \bibnamefont {Yan}}, \ and\ \bibinfo
  {author} {\bibfnamefont {Xianglin}\ \bibnamefont {Ke}},\ }\bibfield  {title}
  {\enquote {\bibinfo {title} {{Exchange-biased topological transverse
  thermoelectric effects in a Kagome ferrimagnet}},}\ }\href {\doibase
  10.1038/s41467-022-28733-7} {\bibfield  {journal} {\bibinfo  {journal}
  {Nature Communications}\ }\textbf {\bibinfo {volume} {13}},\ \bibinfo {pages}
  {1091} (\bibinfo {year} {2022}{\natexlab{a}})}\BibitemShut {NoStop}%
\bibitem [{\citenamefont {Arachchige}\ \emph {et~al.}(2022)\citenamefont
  {Arachchige}, \citenamefont {Meier}, \citenamefont {Marshall}, \citenamefont
  {Matsuoka}, \citenamefont {Xue}, \citenamefont {McGuire}, \citenamefont
  {Hermann}, \citenamefont {Cao},\ and\ \citenamefont
  {Mandrus}}]{Arachchige2022charge}%
  \BibitemOpen
  \bibfield  {author} {\bibinfo {author} {\bibfnamefont {Hasitha W.~Suriya}\
  \bibnamefont {Arachchige}}, \bibinfo {author} {\bibfnamefont {William~R.}\
  \bibnamefont {Meier}}, \bibinfo {author} {\bibfnamefont {Madalynn}\
  \bibnamefont {Marshall}}, \bibinfo {author} {\bibfnamefont {Takahiro}\
  \bibnamefont {Matsuoka}}, \bibinfo {author} {\bibfnamefont {Rui}\
  \bibnamefont {Xue}}, \bibinfo {author} {\bibfnamefont {Michael~A.}\
  \bibnamefont {McGuire}}, \bibinfo {author} {\bibfnamefont {Raphael~P.}\
  \bibnamefont {Hermann}}, \bibinfo {author} {\bibfnamefont {Huibo}\
  \bibnamefont {Cao}}, \ and\ \bibinfo {author} {\bibfnamefont {David}\
  \bibnamefont {Mandrus}},\ }\bibfield  {title} {\enquote {\bibinfo {title}
  {{Charge Density Wave in Kagome Lattice Intermetallic
  ${\mathrm{ScV}}_{6}{\mathrm{Sn}}_{6}$}},}\ }\href {\doibase
  10.1103/PhysRevLett.129.216402} {\bibfield  {journal} {\bibinfo  {journal}
  {Phys. Rev. Lett.}\ }\textbf {\bibinfo {volume} {129}},\ \bibinfo {pages}
  {216402} (\bibinfo {year} {2022})}\BibitemShut {NoStop}%
\bibitem [{\citenamefont {Tan}\ and\ \citenamefont {Yan}(2023)}]{tan2023PRL}%
  \BibitemOpen
  \bibfield  {author} {\bibinfo {author} {\bibfnamefont {Hengxin}\ \bibnamefont
  {Tan}}\ and\ \bibinfo {author} {\bibfnamefont {Binghai}\ \bibnamefont
  {Yan}},\ }\bibfield  {title} {\enquote {\bibinfo {title} {{Abundant Lattice
  Instability in Kagome Metal ${\mathrm{ScV}}_{6}{\mathrm{Sn}}_{6}$}},}\ }\href
  {\doibase 10.1103/PhysRevLett.130.266402} {\bibfield  {journal} {\bibinfo
  {journal} {Phys. Rev. Lett.}\ }\textbf {\bibinfo {volume} {130}},\ \bibinfo
  {pages} {266402} (\bibinfo {year} {2023})}\BibitemShut {NoStop}%
\bibitem [{\citenamefont {Cao}\ \emph {et~al.}(2023)\citenamefont {Cao},
  \citenamefont {Xu}, \citenamefont {Fukui}, \citenamefont {Manjo},
  \citenamefont {Shi}, \citenamefont {Liu}, \citenamefont {Cao},\ and\
  \citenamefont {Song}}]{cao2023competing}%
  \BibitemOpen
  \bibfield  {author} {\bibinfo {author} {\bibfnamefont {Saizheng}\
  \bibnamefont {Cao}}, \bibinfo {author} {\bibfnamefont {Chenchao}\
  \bibnamefont {Xu}}, \bibinfo {author} {\bibfnamefont {Hiroshi}\ \bibnamefont
  {Fukui}}, \bibinfo {author} {\bibfnamefont {Taishun}\ \bibnamefont {Manjo}},
  \bibinfo {author} {\bibfnamefont {Ming}\ \bibnamefont {Shi}}, \bibinfo
  {author} {\bibfnamefont {Yang}\ \bibnamefont {Liu}}, \bibinfo {author}
  {\bibfnamefont {Chao}\ \bibnamefont {Cao}}, \ and\ \bibinfo {author}
  {\bibfnamefont {Yu}~\bibnamefont {Song}},\ }\bibfield  {title} {\enquote
  {\bibinfo {title} {{Competing charge-density wave instabilities in the kagome
  metal ScV$_6$Sn$_6$}},}\ }\href {\doibase 10.1038/s41467-023-43454-1}
  {\bibfield  {journal} {\bibinfo  {journal} {Nature Communications}\ }\textbf
  {\bibinfo {volume} {14}},\ \bibinfo {pages} {7671} (\bibinfo {year}
  {2023})}\BibitemShut {NoStop}%
\bibitem [{\citenamefont {Korshunov}\ \emph {et~al.}(2023)\citenamefont
  {Korshunov}, \citenamefont {Hu}, \citenamefont {Subires}, \citenamefont
  {Jiang}, \citenamefont {Călugăru}, \citenamefont {Feng}, \citenamefont
  {Rajapitamahuni}, \citenamefont {Yi}, \citenamefont {Roychowdhury},
  \citenamefont {Vergniory}, \citenamefont {Strempfer}, \citenamefont
  {Shekhar}, \citenamefont {Vescovo}, \citenamefont {Chernyshov}, \citenamefont
  {Said}, \citenamefont {Bosak}, \citenamefont {Felser}, \citenamefont
  {Bernevig},\ and\ \citenamefont {Blanco-Canosa}}]{Korshunov2023softening}%
  \BibitemOpen
  \bibfield  {author} {\bibinfo {author} {\bibfnamefont {A.}~\bibnamefont
  {Korshunov}}, \bibinfo {author} {\bibfnamefont {H.}~\bibnamefont {Hu}},
  \bibinfo {author} {\bibfnamefont {D.}~\bibnamefont {Subires}}, \bibinfo
  {author} {\bibfnamefont {Y.}~\bibnamefont {Jiang}}, \bibinfo {author}
  {\bibfnamefont {D.}~\bibnamefont {Călugăru}}, \bibinfo {author}
  {\bibfnamefont {X.}~\bibnamefont {Feng}}, \bibinfo {author} {\bibfnamefont
  {A.}~\bibnamefont {Rajapitamahuni}}, \bibinfo {author} {\bibfnamefont
  {C.}~\bibnamefont {Yi}}, \bibinfo {author} {\bibfnamefont {S.}~\bibnamefont
  {Roychowdhury}}, \bibinfo {author} {\bibfnamefont {M.~G.}\ \bibnamefont
  {Vergniory}}, \bibinfo {author} {\bibfnamefont {J.}~\bibnamefont
  {Strempfer}}, \bibinfo {author} {\bibfnamefont {C.}~\bibnamefont {Shekhar}},
  \bibinfo {author} {\bibfnamefont {E.}~\bibnamefont {Vescovo}}, \bibinfo
  {author} {\bibfnamefont {D.}~\bibnamefont {Chernyshov}}, \bibinfo {author}
  {\bibfnamefont {A.~H.}\ \bibnamefont {Said}}, \bibinfo {author}
  {\bibfnamefont {A.}~\bibnamefont {Bosak}}, \bibinfo {author} {\bibfnamefont
  {C.}~\bibnamefont {Felser}}, \bibinfo {author} {\bibfnamefont {B.~Andrei}\
  \bibnamefont {Bernevig}}, \ and\ \bibinfo {author} {\bibfnamefont
  {S.}~\bibnamefont {Blanco-Canosa}},\ }\bibfield  {title} {\enquote {\bibinfo
  {title} {{Softening of a flat phonon mode in the kagome ScV$_6$Sn$_6$}},}\
  }\href {\doibase 10.1038/s41467-023-42186-6} {\bibfield  {journal} {\bibinfo
  {journal} {Nature Communications}\ }\textbf {\bibinfo {volume} {14}},\
  \bibinfo {pages} {6646} (\bibinfo {year} {2023})}\BibitemShut {NoStop}%
\bibitem [{\citenamefont {Cheng}\ \emph {et~al.}(2024)\citenamefont {Cheng},
  \citenamefont {Ren}, \citenamefont {Li}, \citenamefont {Oh}, \citenamefont
  {Tan}, \citenamefont {Pokharel}, \citenamefont {DeStefano}, \citenamefont
  {Rosenberg}, \citenamefont {Guo}, \citenamefont {Zhang}, \citenamefont {Yue},
  \citenamefont {Lee}, \citenamefont {Gorovikov}, \citenamefont {Zonno},
  \citenamefont {Hashimoto}, \citenamefont {Lu}, \citenamefont {Ke},
  \citenamefont {Mazzola}, \citenamefont {Kono}, \citenamefont {Birgeneau},
  \citenamefont {Chu}, \citenamefont {Wilson}, \citenamefont {Wang},
  \citenamefont {Yan}, \citenamefont {Yi},\ and\ \citenamefont
  {Zeljkovic}}]{cheng2023nanoscale}%
  \BibitemOpen
  \bibfield  {author} {\bibinfo {author} {\bibfnamefont {Siyu}\ \bibnamefont
  {Cheng}}, \bibinfo {author} {\bibfnamefont {Zheng}\ \bibnamefont {Ren}},
  \bibinfo {author} {\bibfnamefont {Hong}\ \bibnamefont {Li}}, \bibinfo
  {author} {\bibfnamefont {Jiseop}\ \bibnamefont {Oh}}, \bibinfo {author}
  {\bibfnamefont {Hengxin}\ \bibnamefont {Tan}}, \bibinfo {author}
  {\bibfnamefont {Ganesh}\ \bibnamefont {Pokharel}}, \bibinfo {author}
  {\bibfnamefont {Jonathan~M.}\ \bibnamefont {DeStefano}}, \bibinfo {author}
  {\bibfnamefont {Elliott}\ \bibnamefont {Rosenberg}}, \bibinfo {author}
  {\bibfnamefont {Yucheng}\ \bibnamefont {Guo}}, \bibinfo {author}
  {\bibfnamefont {Yichen}\ \bibnamefont {Zhang}}, \bibinfo {author}
  {\bibfnamefont {Ziqin}\ \bibnamefont {Yue}}, \bibinfo {author} {\bibfnamefont
  {Yongbin}\ \bibnamefont {Lee}}, \bibinfo {author} {\bibfnamefont {Sergey}\
  \bibnamefont {Gorovikov}}, \bibinfo {author} {\bibfnamefont {Marta}\
  \bibnamefont {Zonno}}, \bibinfo {author} {\bibfnamefont {Makoto}\
  \bibnamefont {Hashimoto}}, \bibinfo {author} {\bibfnamefont {Donghui}\
  \bibnamefont {Lu}}, \bibinfo {author} {\bibfnamefont {Liqin}\ \bibnamefont
  {Ke}}, \bibinfo {author} {\bibfnamefont {Federico}\ \bibnamefont {Mazzola}},
  \bibinfo {author} {\bibfnamefont {Junichiro}\ \bibnamefont {Kono}}, \bibinfo
  {author} {\bibfnamefont {R.~J.}\ \bibnamefont {Birgeneau}}, \bibinfo {author}
  {\bibfnamefont {Jiun-Haw}\ \bibnamefont {Chu}}, \bibinfo {author}
  {\bibfnamefont {Stephen~D.}\ \bibnamefont {Wilson}}, \bibinfo {author}
  {\bibfnamefont {Ziqiang}\ \bibnamefont {Wang}}, \bibinfo {author}
  {\bibfnamefont {Binghai}\ \bibnamefont {Yan}}, \bibinfo {author}
  {\bibfnamefont {Ming}\ \bibnamefont {Yi}}, \ and\ \bibinfo {author}
  {\bibfnamefont {Ilija}\ \bibnamefont {Zeljkovic}},\ }\bibfield  {title}
  {\enquote {\bibinfo {title} {{Nanoscale visualization and spectral
  fingerprints of the charge order in ScV$_6$Sn$_6$ distinct from other kagome
  metals}},}\ }\href {\doibase 10.1038/s41535-024-00623-9} {\bibfield
  {journal} {\bibinfo  {journal} {npj Quantum Mater.}\ }\textbf {\bibinfo
  {volume} {9}},\ \bibinfo {pages} {14} (\bibinfo {year} {2024})}\BibitemShut
  {NoStop}%
\bibitem [{\citenamefont {Pokharel}\ \emph {et~al.}(2022)\citenamefont
  {Pokharel}, \citenamefont {Ortiz}, \citenamefont {Chamorro}, \citenamefont
  {Sarte}, \citenamefont {Kautzsch}, \citenamefont {Wu}, \citenamefont {Ruff},\
  and\ \citenamefont {Wilson}}]{Pokharel2022highly}%
  \BibitemOpen
  \bibfield  {author} {\bibinfo {author} {\bibfnamefont {Ganesh}\ \bibnamefont
  {Pokharel}}, \bibinfo {author} {\bibfnamefont {Brenden}\ \bibnamefont
  {Ortiz}}, \bibinfo {author} {\bibfnamefont {Juan}\ \bibnamefont {Chamorro}},
  \bibinfo {author} {\bibfnamefont {Paul}\ \bibnamefont {Sarte}}, \bibinfo
  {author} {\bibfnamefont {Linus}\ \bibnamefont {Kautzsch}}, \bibinfo {author}
  {\bibfnamefont {Guang}\ \bibnamefont {Wu}}, \bibinfo {author} {\bibfnamefont
  {Jacob}\ \bibnamefont {Ruff}}, \ and\ \bibinfo {author} {\bibfnamefont
  {Stephen~D.}\ \bibnamefont {Wilson}},\ }\bibfield  {title} {\enquote
  {\bibinfo {title} {{Highly anisotropic magnetism in the vanadium-based kagome
  metal ${\mathrm{TbV}}_{6}{\mathrm{Sn}}_{6}$}},}\ }\href {\doibase
  10.1103/PhysRevMaterials.6.104202} {\bibfield  {journal} {\bibinfo  {journal}
  {Phys. Rev. Mater.}\ }\textbf {\bibinfo {volume} {6}},\ \bibinfo {pages}
  {104202} (\bibinfo {year} {2022})}\BibitemShut {NoStop}%
\bibitem [{\citenamefont {Rosenberg}\ \emph {et~al.}(2022)\citenamefont
  {Rosenberg}, \citenamefont {DeStefano}, \citenamefont {Guo}, \citenamefont
  {Oh}, \citenamefont {Hashimoto}, \citenamefont {Lu}, \citenamefont
  {Birgeneau}, \citenamefont {Lee}, \citenamefont {Ke}, \citenamefont {Yi},\
  and\ \citenamefont {Chu}}]{Rosenberg2022uniaxial}%
  \BibitemOpen
  \bibfield  {author} {\bibinfo {author} {\bibfnamefont {Elliott}\ \bibnamefont
  {Rosenberg}}, \bibinfo {author} {\bibfnamefont {Jonathan~M.}\ \bibnamefont
  {DeStefano}}, \bibinfo {author} {\bibfnamefont {Yucheng}\ \bibnamefont
  {Guo}}, \bibinfo {author} {\bibfnamefont {Ji~Seop}\ \bibnamefont {Oh}},
  \bibinfo {author} {\bibfnamefont {Makoto}\ \bibnamefont {Hashimoto}},
  \bibinfo {author} {\bibfnamefont {Donghui}\ \bibnamefont {Lu}}, \bibinfo
  {author} {\bibfnamefont {Robert~J.}\ \bibnamefont {Birgeneau}}, \bibinfo
  {author} {\bibfnamefont {Yongbin}\ \bibnamefont {Lee}}, \bibinfo {author}
  {\bibfnamefont {Liqin}\ \bibnamefont {Ke}}, \bibinfo {author} {\bibfnamefont
  {Ming}\ \bibnamefont {Yi}}, \ and\ \bibinfo {author} {\bibfnamefont
  {Jiun-Haw}\ \bibnamefont {Chu}},\ }\bibfield  {title} {\enquote {\bibinfo
  {title} {{Uniaxial ferromagnetism in the kagome metal
  ${\mathrm{TbV}}_{6}{\mathrm{Sn}}_{6}$}},}\ }\href {\doibase
  10.1103/PhysRevB.106.115139} {\bibfield  {journal} {\bibinfo  {journal}
  {Phys. Rev. B}\ }\textbf {\bibinfo {volume} {106}},\ \bibinfo {pages}
  {115139} (\bibinfo {year} {2022})}\BibitemShut {NoStop}%
\bibitem [{\citenamefont {Zhang}\ \emph
  {et~al.}(2022{\natexlab{b}})\citenamefont {Zhang}, \citenamefont {Liu},
  \citenamefont {Cui}, \citenamefont {Guo}, \citenamefont {Wang}, \citenamefont
  {Shi}, \citenamefont {Zhang}, \citenamefont {Wang}, \citenamefont {Dong},
  \citenamefont {Sun}, \citenamefont {Dun},\ and\ \citenamefont
  {Cheng}}]{zhang2022electronic}%
  \BibitemOpen
  \bibfield  {author} {\bibinfo {author} {\bibfnamefont {Xiaoxiao}\
  \bibnamefont {Zhang}}, \bibinfo {author} {\bibfnamefont {Ziyi}\ \bibnamefont
  {Liu}}, \bibinfo {author} {\bibfnamefont {Qi}~\bibnamefont {Cui}}, \bibinfo
  {author} {\bibfnamefont {Qi}~\bibnamefont {Guo}}, \bibinfo {author}
  {\bibfnamefont {Ningning}\ \bibnamefont {Wang}}, \bibinfo {author}
  {\bibfnamefont {Lifen}\ \bibnamefont {Shi}}, \bibinfo {author} {\bibfnamefont
  {Hua}\ \bibnamefont {Zhang}}, \bibinfo {author} {\bibfnamefont {Weihua}\
  \bibnamefont {Wang}}, \bibinfo {author} {\bibfnamefont {Xiaoli}\ \bibnamefont
  {Dong}}, \bibinfo {author} {\bibfnamefont {Jianping}\ \bibnamefont {Sun}},
  \bibinfo {author} {\bibfnamefont {Zhiling}\ \bibnamefont {Dun}}, \ and\
  \bibinfo {author} {\bibfnamefont {Jinguang}\ \bibnamefont {Cheng}},\
  }\bibfield  {title} {\enquote {\bibinfo {title} {{Electronic and magnetic
  properties of intermetallic kagome magnets
  $R{\mathrm{V}}_{6}{\mathrm{Sn}}_{6}(R=\mathrm{Tb}\text{\ensuremath{-}}\mathrm{Tm})$}},}\
  }\href {\doibase 10.1103/PhysRevMaterials.6.105001} {\bibfield  {journal}
  {\bibinfo  {journal} {Phys. Rev. Mater.}\ }\textbf {\bibinfo {volume} {6}},\
  \bibinfo {pages} {105001} (\bibinfo {year} {2022}{\natexlab{b}})}\BibitemShut
  {NoStop}%
\bibitem [{\citenamefont {Lee}\ and\ \citenamefont
  {Mun}(2022)}]{lee2022anisotropic}%
  \BibitemOpen
  \bibfield  {author} {\bibinfo {author} {\bibfnamefont {Jeonghun}\
  \bibnamefont {Lee}}\ and\ \bibinfo {author} {\bibfnamefont {Eundeok}\
  \bibnamefont {Mun}},\ }\bibfield  {title} {\enquote {\bibinfo {title}
  {{Anisotropic magnetic property of single crystals
  $R{\mathrm{V}}_{6}{\mathrm{Sn}}_{6}$ $(R=\mathrm{Y},
  \mathrm{Gd}\text{\ensuremath{-}}\mathrm{Tm}, \mathrm{Lu})$}},}\ }\href
  {\doibase 10.1103/PhysRevMaterials.6.083401} {\bibfield  {journal} {\bibinfo
  {journal} {Phys. Rev. Mater.}\ }\textbf {\bibinfo {volume} {6}},\ \bibinfo
  {pages} {083401} (\bibinfo {year} {2022})}\BibitemShut {NoStop}%
\bibitem [{\citenamefont {Peng}\ \emph {et~al.}(2021)\citenamefont {Peng},
  \citenamefont {Han}, \citenamefont {Pokharel}, \citenamefont {Shen},
  \citenamefont {Li}, \citenamefont {Hashimoto}, \citenamefont {Lu},
  \citenamefont {Ortiz}, \citenamefont {Luo}, \citenamefont {Li}, \citenamefont
  {Guo}, \citenamefont {Wang}, \citenamefont {Cui}, \citenamefont {Sun},
  \citenamefont {Qiao}, \citenamefont {Wilson},\ and\ \citenamefont
  {He}}]{peng2021realizing}%
  \BibitemOpen
  \bibfield  {author} {\bibinfo {author} {\bibfnamefont {Shuting}\ \bibnamefont
  {Peng}}, \bibinfo {author} {\bibfnamefont {Yulei}\ \bibnamefont {Han}},
  \bibinfo {author} {\bibfnamefont {Ganesh}\ \bibnamefont {Pokharel}}, \bibinfo
  {author} {\bibfnamefont {Jianchang}\ \bibnamefont {Shen}}, \bibinfo {author}
  {\bibfnamefont {Zeyu}\ \bibnamefont {Li}}, \bibinfo {author} {\bibfnamefont
  {Makoto}\ \bibnamefont {Hashimoto}}, \bibinfo {author} {\bibfnamefont
  {Donghui}\ \bibnamefont {Lu}}, \bibinfo {author} {\bibfnamefont {Brenden~R.}\
  \bibnamefont {Ortiz}}, \bibinfo {author} {\bibfnamefont {Yang}\ \bibnamefont
  {Luo}}, \bibinfo {author} {\bibfnamefont {Houchen}\ \bibnamefont {Li}},
  \bibinfo {author} {\bibfnamefont {Mingyao}\ \bibnamefont {Guo}}, \bibinfo
  {author} {\bibfnamefont {Bingqian}\ \bibnamefont {Wang}}, \bibinfo {author}
  {\bibfnamefont {Shengtao}\ \bibnamefont {Cui}}, \bibinfo {author}
  {\bibfnamefont {Zhe}\ \bibnamefont {Sun}}, \bibinfo {author} {\bibfnamefont
  {Zhenhua}\ \bibnamefont {Qiao}}, \bibinfo {author} {\bibfnamefont
  {Stephen~D.}\ \bibnamefont {Wilson}}, \ and\ \bibinfo {author} {\bibfnamefont
  {Junfeng}\ \bibnamefont {He}},\ }\bibfield  {title} {\enquote {\bibinfo
  {title} {{Realizing Kagome Band Structure in Two-Dimensional Kagome Surface
  States of $R{\mathrm{V}}_{6}{\mathrm{Sn}}_{6}$ ($R$ = Gd, Ho)}},}\ }\href
  {\doibase 10.1103/PhysRevLett.127.266401} {\bibfield  {journal} {\bibinfo
  {journal} {Phys. Rev. Lett.}\ }\textbf {\bibinfo {volume} {127}},\ \bibinfo
  {pages} {266401} (\bibinfo {year} {2021})}\BibitemShut {NoStop}%
\bibitem [{\citenamefont {Hu}\ \emph {et~al.}(2022)\citenamefont {Hu},
  \citenamefont {Wu}, \citenamefont {Yang}, \citenamefont {Gao}, \citenamefont
  {Plumb}, \citenamefont {Schnyder}, \citenamefont {Xie}, \citenamefont {Ma},\
  and\ \citenamefont {Shi}}]{hu2022tunable}%
  \BibitemOpen
  \bibfield  {author} {\bibinfo {author} {\bibfnamefont {Yong}\ \bibnamefont
  {Hu}}, \bibinfo {author} {\bibfnamefont {Xianxin}\ \bibnamefont {Wu}},
  \bibinfo {author} {\bibfnamefont {Yongqi}\ \bibnamefont {Yang}}, \bibinfo
  {author} {\bibfnamefont {Shunye}\ \bibnamefont {Gao}}, \bibinfo {author}
  {\bibfnamefont {Nicholas~C.}\ \bibnamefont {Plumb}}, \bibinfo {author}
  {\bibfnamefont {Andreas~P.}\ \bibnamefont {Schnyder}}, \bibinfo {author}
  {\bibfnamefont {Weiwei}\ \bibnamefont {Xie}}, \bibinfo {author}
  {\bibfnamefont {Junzhang}\ \bibnamefont {Ma}}, \ and\ \bibinfo {author}
  {\bibfnamefont {Ming}\ \bibnamefont {Shi}},\ }\bibfield  {title} {\enquote
  {\bibinfo {title} {{Tunable topological Dirac surface states and van Hove
  singularities in kagome metal GdV$_6$Sn$_6$}},}\ }\href {\doibase
  10.1126/sciadv.add2024} {\bibfield  {journal} {\bibinfo  {journal} {Science
  Advances}\ }\textbf {\bibinfo {volume} {8}},\ \bibinfo {pages} {eadd2024}
  (\bibinfo {year} {2022})}\BibitemShut {NoStop}%
\bibitem [{\citenamefont {Sante}\ \emph {et~al.}(2023)\citenamefont {Sante},
  \citenamefont {Bigi}, \citenamefont {Eck}, \citenamefont {Enzner},
  \citenamefont {Consiglio}, \citenamefont {Pokharel}, \citenamefont {Carrara},
  \citenamefont {Orgiani}, \citenamefont {Polewczyk}, \citenamefont {Fujii},
  \citenamefont {King}, \citenamefont {Vobornik}, \citenamefont {Rossi},
  \citenamefont {Zeljkovic}, \citenamefont {Wilson}, \citenamefont {Thomale},
  \citenamefont {Sangiovanni}, \citenamefont {Panaccione},\ and\ \citenamefont
  {Mazzola}}]{Sante2023flat}%
  \BibitemOpen
  \bibfield  {author} {\bibinfo {author} {\bibfnamefont {Domenico~Di}\
  \bibnamefont {Sante}}, \bibinfo {author} {\bibfnamefont {Chiara}\
  \bibnamefont {Bigi}}, \bibinfo {author} {\bibfnamefont {Philipp}\
  \bibnamefont {Eck}}, \bibinfo {author} {\bibfnamefont {Stefan}\ \bibnamefont
  {Enzner}}, \bibinfo {author} {\bibfnamefont {Armando}\ \bibnamefont
  {Consiglio}}, \bibinfo {author} {\bibfnamefont {Ganesh}\ \bibnamefont
  {Pokharel}}, \bibinfo {author} {\bibfnamefont {Pietro}\ \bibnamefont
  {Carrara}}, \bibinfo {author} {\bibfnamefont {Pasquale}\ \bibnamefont
  {Orgiani}}, \bibinfo {author} {\bibfnamefont {Vincent}\ \bibnamefont
  {Polewczyk}}, \bibinfo {author} {\bibfnamefont {Jun}\ \bibnamefont {Fujii}},
  \bibinfo {author} {\bibfnamefont {Phil D.~C.}\ \bibnamefont {King}}, \bibinfo
  {author} {\bibfnamefont {Ivana}\ \bibnamefont {Vobornik}}, \bibinfo {author}
  {\bibfnamefont {Giorgio}\ \bibnamefont {Rossi}}, \bibinfo {author}
  {\bibfnamefont {Ilija}\ \bibnamefont {Zeljkovic}}, \bibinfo {author}
  {\bibfnamefont {Stephen~D.}\ \bibnamefont {Wilson}}, \bibinfo {author}
  {\bibfnamefont {Ronny}\ \bibnamefont {Thomale}}, \bibinfo {author}
  {\bibfnamefont {Giorgio}\ \bibnamefont {Sangiovanni}}, \bibinfo {author}
  {\bibfnamefont {Giancarlo}\ \bibnamefont {Panaccione}}, \ and\ \bibinfo
  {author} {\bibfnamefont {Federico}\ \bibnamefont {Mazzola}},\ }\bibfield
  {title} {\enquote {\bibinfo {title} {Flat band separation and robust spin
  berry curvature in bilayer kagome metals},}\ }\href {\doibase
  10.1038/s41567-023-02053-z} {\bibfield  {journal} {\bibinfo  {journal}
  {Nature Physics}\ }\textbf {\bibinfo {volume} {19}},\ \bibinfo {pages}
  {1135–1142} (\bibinfo {year} {2023})}\BibitemShut {NoStop}%
\bibitem [{\citenamefont {Pokharel}\ \emph {et~al.}(2021)\citenamefont
  {Pokharel}, \citenamefont {Teicher}, \citenamefont {Ortiz}, \citenamefont
  {Sarte}, \citenamefont {Wu}, \citenamefont {Peng}, \citenamefont {He},
  \citenamefont {Seshadri},\ and\ \citenamefont
  {Wilson}}]{Pokharel2021electronic}%
  \BibitemOpen
  \bibfield  {author} {\bibinfo {author} {\bibfnamefont {Ganesh}\ \bibnamefont
  {Pokharel}}, \bibinfo {author} {\bibfnamefont {Samuel M.~L.}\ \bibnamefont
  {Teicher}}, \bibinfo {author} {\bibfnamefont {Brenden~R.}\ \bibnamefont
  {Ortiz}}, \bibinfo {author} {\bibfnamefont {Paul~M.}\ \bibnamefont {Sarte}},
  \bibinfo {author} {\bibfnamefont {Guang}\ \bibnamefont {Wu}}, \bibinfo
  {author} {\bibfnamefont {Shuting}\ \bibnamefont {Peng}}, \bibinfo {author}
  {\bibfnamefont {Junfeng}\ \bibnamefont {He}}, \bibinfo {author}
  {\bibfnamefont {Ram}\ \bibnamefont {Seshadri}}, \ and\ \bibinfo {author}
  {\bibfnamefont {Stephen~D.}\ \bibnamefont {Wilson}},\ }\bibfield  {title}
  {\enquote {\bibinfo {title} {{Electronic properties of the topological kagome
  metals ${\mathrm{YV}}_{6}{\mathrm{Sn}}_{6}$ and
  ${\mathrm{GdV}}_{6}{\mathrm{Sn}}_{6}$}},}\ }\href {\doibase
  10.1103/PhysRevB.104.235139} {\bibfield  {journal} {\bibinfo  {journal}
  {Phys. Rev. B}\ }\textbf {\bibinfo {volume} {104}},\ \bibinfo {pages}
  {235139} (\bibinfo {year} {2021})}\BibitemShut {NoStop}%
\bibitem [{\citenamefont {Ding}\ \emph {et~al.}(2023)\citenamefont {Ding},
  \citenamefont {Zhao}, \citenamefont {Tao}, \citenamefont {Huang},
  \citenamefont {Jiang}, \citenamefont {Yang}, \citenamefont {Cho},
  \citenamefont {Liu}, \citenamefont {Liu}, \citenamefont {Guo}, \citenamefont
  {Liu}, \citenamefont {Liu},\ and\ \citenamefont {Shen}}]{ding2023kagome}%
  \BibitemOpen
  \bibfield  {author} {\bibinfo {author} {\bibfnamefont {Jianyang}\
  \bibnamefont {Ding}}, \bibinfo {author} {\bibfnamefont {Ningning}\
  \bibnamefont {Zhao}}, \bibinfo {author} {\bibfnamefont {Zicheng}\
  \bibnamefont {Tao}}, \bibinfo {author} {\bibfnamefont {Zhe}\ \bibnamefont
  {Huang}}, \bibinfo {author} {\bibfnamefont {Zhicheng}\ \bibnamefont {Jiang}},
  \bibinfo {author} {\bibfnamefont {Yichen}\ \bibnamefont {Yang}}, \bibinfo
  {author} {\bibfnamefont {Soohyun}\ \bibnamefont {Cho}}, \bibinfo {author}
  {\bibfnamefont {Zhengtai}\ \bibnamefont {Liu}}, \bibinfo {author}
  {\bibfnamefont {Jishan}\ \bibnamefont {Liu}}, \bibinfo {author}
  {\bibfnamefont {Yanfeng}\ \bibnamefont {Guo}}, \bibinfo {author}
  {\bibfnamefont {Kai}\ \bibnamefont {Liu}}, \bibinfo {author} {\bibfnamefont
  {Zhonghao}\ \bibnamefont {Liu}}, \ and\ \bibinfo {author} {\bibfnamefont
  {Dawei}\ \bibnamefont {Shen}},\ }\bibfield  {title} {\enquote {\bibinfo
  {title} {Kagome surface states and weak electronic correlation in
  vanadium-kagome metals},}\ }\href {\doibase 10.1088/1361-648X/ace2a2}
  {\bibfield  {journal} {\bibinfo  {journal} {Journal of Physics: Condensed
  Matter}\ }\textbf {\bibinfo {volume} {35}},\ \bibinfo {pages} {405502}
  (\bibinfo {year} {2023})}\BibitemShut {NoStop}%
\bibitem [{SM()}]{SM}%
  \BibitemOpen
  \href@noop {} {}\bibinfo {note} {See the Supplemental Material for methods
  employed in our calculations and extended results and discussions about the
  bulk, surfaces, and thin films of TbV$_6$Sn$_6$ and YV$_6$Sn$_6$, and
  properties of magnetic kagome surfaces (Mn-doped TbV$_6$Sn$_6$,
  TbMn$_6$Sn$_6$ and V-doped TbMn$_6$Sn$_6$). Refs.
  \cite{PRB54p11169,VASP,PRL77p3865,vaspkit,phonopy,tan2023PRL,hu2022tunable,hu2023phonon,Arachchige2022charge,Ishikawa2021GdV6Sn6,Rosenberg2022uniaxial,Romaka2011Peculiarities}
  are included.}\BibitemShut {Stop}%
\bibitem [{not()}]{note1}%
  \BibitemOpen
  \href@noop {} {}\bibinfo {note} {After structural relaxation, the V-kagome
  sublattice goes below the closely associated Sn sublattice in the V$_3$Sn
  termination.}\BibitemShut {Stop}%
\bibitem [{\citenamefont {Hu}\ \emph {et~al.}(2023)\citenamefont {Hu},
  \citenamefont {Jiang}, \citenamefont {Călugăru}, \citenamefont {Feng},
  \citenamefont {Subires}, \citenamefont {Vergniory}, \citenamefont {Felser},
  \citenamefont {Blanco-Canosa},\ and\ \citenamefont
  {Bernevig}}]{hu2023kagome}%
  \BibitemOpen
  \bibfield  {author} {\bibinfo {author} {\bibfnamefont {Haoyu}\ \bibnamefont
  {Hu}}, \bibinfo {author} {\bibfnamefont {Yi}~\bibnamefont {Jiang}}, \bibinfo
  {author} {\bibfnamefont {Dumitru}\ \bibnamefont {Călugăru}}, \bibinfo
  {author} {\bibfnamefont {Xiaolong}\ \bibnamefont {Feng}}, \bibinfo {author}
  {\bibfnamefont {David}\ \bibnamefont {Subires}}, \bibinfo {author}
  {\bibfnamefont {Maia~G.}\ \bibnamefont {Vergniory}}, \bibinfo {author}
  {\bibfnamefont {Claudia}\ \bibnamefont {Felser}}, \bibinfo {author}
  {\bibfnamefont {Santiago}\ \bibnamefont {Blanco-Canosa}}, \ and\ \bibinfo
  {author} {\bibfnamefont {B.~Andrei}\ \bibnamefont {Bernevig}},\ }\bibfield
  {title} {\enquote {\bibinfo {title} {{Kagome Materials I: SG 191,
  ScV$_6$Sn$_6$. Flat Phonon Soft Modes and Unconventional CDW Formation:
  Microscopic and Effective Theory}},}\ }\href
  {https://arxiv.org/abs/2305.15469} {\bibfield  {journal} {\bibinfo  {journal}
  {arXiv:2305.15469}\ } (\bibinfo {year} {2023})}\BibitemShut {NoStop}%
\bibitem [{\citenamefont {Shannon}\ and\ \citenamefont
  {Prewitt}(1969)}]{shannon1969effective}%
  \BibitemOpen
  \bibfield  {author} {\bibinfo {author} {\bibfnamefont {R.~D.}\ \bibnamefont
  {Shannon}}\ and\ \bibinfo {author} {\bibfnamefont {C.~T.}\ \bibnamefont
  {Prewitt}},\ }\bibfield  {title} {\enquote {\bibinfo {title} {Effective ionic
  radii in oxides and fluorides},}\ }\href {\doibase 10.1107/S0567740869003220}
  {\bibfield  {journal} {\bibinfo  {journal} {Acta Crystallographica Section
  B}\ }\textbf {\bibinfo {volume} {25}},\ \bibinfo {pages} {925--946} (\bibinfo
  {year} {1969})}\BibitemShut {NoStop}%
\bibitem [{\citenamefont {Wu}\ \emph {et~al.}(2024)\citenamefont {Wu},
  \citenamefont {Klemm}, \citenamefont {Shah}, \citenamefont {Ritz},
  \citenamefont {Duan}, \citenamefont {Teng}, \citenamefont {Gao},
  \citenamefont {Ye}, \citenamefont {Matsuda}, \citenamefont {Li},
  \citenamefont {Xu}, \citenamefont {Yi}, \citenamefont {Birol}, \citenamefont
  {Dai},\ and\ \citenamefont {Blumberg}}]{wu2024symmetry}%
  \BibitemOpen
  \bibfield  {author} {\bibinfo {author} {\bibfnamefont {Shangfei}\
  \bibnamefont {Wu}}, \bibinfo {author} {\bibfnamefont {Mason~L.}\ \bibnamefont
  {Klemm}}, \bibinfo {author} {\bibfnamefont {Jay}\ \bibnamefont {Shah}},
  \bibinfo {author} {\bibfnamefont {Ethan~T.}\ \bibnamefont {Ritz}}, \bibinfo
  {author} {\bibfnamefont {Chunruo}\ \bibnamefont {Duan}}, \bibinfo {author}
  {\bibfnamefont {Xiaokun}\ \bibnamefont {Teng}}, \bibinfo {author}
  {\bibfnamefont {Bin}\ \bibnamefont {Gao}}, \bibinfo {author} {\bibfnamefont
  {Feng}\ \bibnamefont {Ye}}, \bibinfo {author} {\bibfnamefont {Masaaki}\
  \bibnamefont {Matsuda}}, \bibinfo {author} {\bibfnamefont {Fankang}\
  \bibnamefont {Li}}, \bibinfo {author} {\bibfnamefont {Xianghan}\ \bibnamefont
  {Xu}}, \bibinfo {author} {\bibfnamefont {Ming}\ \bibnamefont {Yi}}, \bibinfo
  {author} {\bibfnamefont {Turan}\ \bibnamefont {Birol}}, \bibinfo {author}
  {\bibfnamefont {Pengcheng}\ \bibnamefont {Dai}}, \ and\ \bibinfo {author}
  {\bibfnamefont {Girsh}\ \bibnamefont {Blumberg}},\ }\bibfield  {title}
  {\enquote {\bibinfo {title} {{Symmetry Breaking and Ascending in the Magnetic
  Kagome Metal FeGe}},}\ }\href {\doibase 10.1103/PhysRevX.14.011043}
  {\bibfield  {journal} {\bibinfo  {journal} {Phys. Rev. X}\ }\textbf {\bibinfo
  {volume} {14}},\ \bibinfo {pages} {011043} (\bibinfo {year}
  {2024})}\BibitemShut {NoStop}%
\bibitem [{\citenamefont {Lee}(1955)}]{Lee1955Magnetostriction}%
  \BibitemOpen
  \bibfield  {author} {\bibinfo {author} {\bibfnamefont {E.~W.}\ \bibnamefont
  {Lee}},\ }\bibfield  {title} {\enquote {\bibinfo {title} {Magnetostriction
  and magnetomechanical effects},}\ }\href {\doibase
  10.1088/0034-4885/18/1/305} {\bibfield  {journal} {\bibinfo  {journal}
  {Reports on Progress in Physics}\ }\textbf {\bibinfo {volume} {18}},\
  \bibinfo {pages} {184} (\bibinfo {year} {1955})}\BibitemShut {NoStop}%
\bibitem [{\citenamefont {Kresse}\ and\ \citenamefont
  {Furthm{\"u}ller}(1996{\natexlab{a}})}]{PRB54p11169}%
  \BibitemOpen
  \bibfield  {author} {\bibinfo {author} {\bibfnamefont {G.}~\bibnamefont
  {Kresse}}\ and\ \bibinfo {author} {\bibfnamefont {J.}~\bibnamefont
  {Furthm{\"u}ller}},\ }\bibfield  {title} {\enquote {\bibinfo {title}
  {Efficient iterative schemes for ab initio total-energy calculations using a
  plane-wave basis set},}\ }\href {\doibase 10.1103/PhysRevB.54.11169}
  {\bibfield  {journal} {\bibinfo  {journal} {Phys. Rev. B}\ }\textbf {\bibinfo
  {volume} {54}},\ \bibinfo {pages} {11169} (\bibinfo {year}
  {1996}{\natexlab{a}})}\BibitemShut {NoStop}%
\bibitem [{\citenamefont {Kresse}\ and\ \citenamefont
  {Furthm{\"u}ller}(1996{\natexlab{b}})}]{VASP}%
  \BibitemOpen
  \bibfield  {author} {\bibinfo {author} {\bibfnamefont {Georg}\ \bibnamefont
  {Kresse}}\ and\ \bibinfo {author} {\bibfnamefont {J{\"u}rgen}\ \bibnamefont
  {Furthm{\"u}ller}},\ }\bibfield  {title} {\enquote {\bibinfo {title}
  {Efficiency of ab-initio total energy calculations for metals and
  semiconductors using a plane-wave basis set},}\ }\href {\doibase
  https://doi.org/10.1016/0927-0256(96)00008-0} {\bibfield  {journal} {\bibinfo
   {journal} {Comput. Mater. Sci.}\ }\textbf {\bibinfo {volume} {6}},\ \bibinfo
  {pages} {15--50} (\bibinfo {year} {1996}{\natexlab{b}})}\BibitemShut
  {NoStop}%
\bibitem [{\citenamefont {Perdew}\ \emph {et~al.}(1996)\citenamefont {Perdew},
  \citenamefont {Burke},\ and\ \citenamefont {Ernzerhof}}]{PRL77p3865}%
  \BibitemOpen
  \bibfield  {author} {\bibinfo {author} {\bibfnamefont {John~P.}\ \bibnamefont
  {Perdew}}, \bibinfo {author} {\bibfnamefont {Kieron}\ \bibnamefont {Burke}},
  \ and\ \bibinfo {author} {\bibfnamefont {Matthias}\ \bibnamefont
  {Ernzerhof}},\ }\bibfield  {title} {\enquote {\bibinfo {title} {Generalized
  gradient approximation made simple},}\ }\href {\doibase
  10.1103/PhysRevLett.77.3865} {\bibfield  {journal} {\bibinfo  {journal}
  {Phys. Rev. Lett.}\ }\textbf {\bibinfo {volume} {77}},\ \bibinfo {pages}
  {3865} (\bibinfo {year} {1996})}\BibitemShut {NoStop}%
\bibitem [{\citenamefont {Wang}\ \emph {et~al.}(2021)\citenamefont {Wang},
  \citenamefont {Xu}, \citenamefont {Liu}, \citenamefont {Tang},\ and\
  \citenamefont {Geng}}]{vaspkit}%
  \BibitemOpen
  \bibfield  {author} {\bibinfo {author} {\bibfnamefont {Vei}\ \bibnamefont
  {Wang}}, \bibinfo {author} {\bibfnamefont {Nan}\ \bibnamefont {Xu}}, \bibinfo
  {author} {\bibfnamefont {Jin-Cheng}\ \bibnamefont {Liu}}, \bibinfo {author}
  {\bibfnamefont {Gang}\ \bibnamefont {Tang}}, \ and\ \bibinfo {author}
  {\bibfnamefont {Wen-Tong}\ \bibnamefont {Geng}},\ }\bibfield  {title}
  {\enquote {\bibinfo {title} {{VASPKIT: A user-friendly interface facilitating
  high-throughput computing and analysis using VASP code}},}\ }\href {\doibase
  https://doi.org/10.1016/j.cpc.2021.108033} {\bibfield  {journal} {\bibinfo
  {journal} {Computer Physics Communications}\ }\textbf {\bibinfo {volume}
  {267}},\ \bibinfo {pages} {108033} (\bibinfo {year} {2021})}\BibitemShut
  {NoStop}%
\bibitem [{\citenamefont {Togo}\ and\ \citenamefont {Tanaka}(2015)}]{phonopy}%
  \BibitemOpen
  \bibfield  {author} {\bibinfo {author} {\bibfnamefont {A}~\bibnamefont
  {Togo}}\ and\ \bibinfo {author} {\bibfnamefont {I}~\bibnamefont {Tanaka}},\
  }\bibfield  {title} {\enquote {\bibinfo {title} {First principles phonon
  calculations in materials science},}\ }\href {\doibase
  https://doi.org/10.1016/j.scriptamat.2015.07.021} {\bibfield  {journal}
  {\bibinfo  {journal} {Scr. Mater.}\ }\textbf {\bibinfo {volume} {108}},\
  \bibinfo {pages} {1--5} (\bibinfo {year} {2015})}\BibitemShut {NoStop}%
\bibitem [{\citenamefont {Hu}\ \emph {et~al.}(2024)\citenamefont {Hu},
  \citenamefont {Ma}, \citenamefont {Li}, \citenamefont {Jiang}, \citenamefont
  {Gawryluk}, \citenamefont {Hu}, \citenamefont {Teyssier}, \citenamefont
  {Multian}, \citenamefont {Yin}, \citenamefont {Xu} \emph
  {et~al.}}]{hu2023phonon}%
  \BibitemOpen
  \bibfield  {author} {\bibinfo {author} {\bibfnamefont {Yong}\ \bibnamefont
  {Hu}}, \bibinfo {author} {\bibfnamefont {Junzhang}\ \bibnamefont {Ma}},
  \bibinfo {author} {\bibfnamefont {Yinxiang}\ \bibnamefont {Li}}, \bibinfo
  {author} {\bibfnamefont {Yuxiao}\ \bibnamefont {Jiang}}, \bibinfo {author}
  {\bibfnamefont {Dariusz~Jakub}\ \bibnamefont {Gawryluk}}, \bibinfo {author}
  {\bibfnamefont {Tianchen}\ \bibnamefont {Hu}}, \bibinfo {author}
  {\bibfnamefont {J{\'e}r{\'e}mie}\ \bibnamefont {Teyssier}}, \bibinfo {author}
  {\bibfnamefont {Volodymyr}\ \bibnamefont {Multian}}, \bibinfo {author}
  {\bibfnamefont {Zhouyi}\ \bibnamefont {Yin}}, \bibinfo {author}
  {\bibfnamefont {Shuxiang}\ \bibnamefont {Xu}},  \emph {et~al.},\ }\bibfield
  {title} {\enquote {\bibinfo {title} {{Phonon promoted charge density wave in
  topological kagome metal ScV$_6$Sn$_6$}},}\ }\href {\doibase
  10.1038/s41467-024-45859-y} {\bibfield  {journal} {\bibinfo  {journal}
  {Nature Communications}\ }\textbf {\bibinfo {volume} {15}},\ \bibinfo {pages}
  {1658} (\bibinfo {year} {2024})}\BibitemShut {NoStop}%
\bibitem [{\citenamefont {Ishikawa}\ \emph {et~al.}(2021)\citenamefont
  {Ishikawa}, \citenamefont {Yajima}, \citenamefont {Kawamura}, \citenamefont
  {Mitamura},\ and\ \citenamefont {Kindo}}]{Ishikawa2021GdV6Sn6}%
  \BibitemOpen
  \bibfield  {author} {\bibinfo {author} {\bibfnamefont {Hajime}\ \bibnamefont
  {Ishikawa}}, \bibinfo {author} {\bibfnamefont {Takeshi}\ \bibnamefont
  {Yajima}}, \bibinfo {author} {\bibfnamefont {Mitsuaki}\ \bibnamefont
  {Kawamura}}, \bibinfo {author} {\bibfnamefont {Hiroyuki}\ \bibnamefont
  {Mitamura}}, \ and\ \bibinfo {author} {\bibfnamefont {Koichi}\ \bibnamefont
  {Kindo}},\ }\bibfield  {title} {\enquote {\bibinfo {title} {{GdV$_6$Sn$_6$: A
  Multi-carrier Metal with Non-magnetic 3d-electron Kagome Bands and
  4f-electron Magnetism}},}\ }\href {\doibase 10.7566/JPSJ.90.124704}
  {\bibfield  {journal} {\bibinfo  {journal} {Journal of the Physical Society
  of Japan}\ }\textbf {\bibinfo {volume} {90}},\ \bibinfo {pages} {124704}
  (\bibinfo {year} {2021})}\BibitemShut {NoStop}%
\bibitem [{\citenamefont {Romaka}\ \emph {et~al.}(2011)\citenamefont {Romaka},
  \citenamefont {Stadnyk}, \citenamefont {Romaka}, \citenamefont {Demchenko},
  \citenamefont {Stadnyshyn},\ and\ \citenamefont
  {Konyk}}]{Romaka2011Peculiarities}%
  \BibitemOpen
  \bibfield  {author} {\bibinfo {author} {\bibfnamefont {L.}~\bibnamefont
  {Romaka}}, \bibinfo {author} {\bibfnamefont {Yu.}\ \bibnamefont {Stadnyk}},
  \bibinfo {author} {\bibfnamefont {V.V.}\ \bibnamefont {Romaka}}, \bibinfo
  {author} {\bibfnamefont {P.}~\bibnamefont {Demchenko}}, \bibinfo {author}
  {\bibfnamefont {M.}~\bibnamefont {Stadnyshyn}}, \ and\ \bibinfo {author}
  {\bibfnamefont {M.}~\bibnamefont {Konyk}},\ }\bibfield  {title} {\enquote
  {\bibinfo {title} {{Peculiarities of component interaction in \{Gd,
  Er\}–V–Sn Ternary systems at 870 K and crystal structure of
  $R$V$_6$Sn$_6$ stannides}},}\ }\href {\doibase
  https://doi.org/10.1016/j.jallcom.2011.06.095} {\bibfield  {journal}
  {\bibinfo  {journal} {Journal of Alloys and Compounds}\ }\textbf {\bibinfo
  {volume} {509}},\ \bibinfo {pages} {8862--8869} (\bibinfo {year}
  {2011})}\BibitemShut {NoStop}%
\end{thebibliography}
%

\end{document}